%% file: paper.tex
\documentclass[fleqn,usenatbib]{mnras}
\usepackage[T1]{fontenc}
\usepackage{ae,aecompl}

\usepackage{aas_macros}
\usepackage{verbatim, amsmath, amsfonts, amssymb}
\usepackage[usenames,dvipsnames,table]{xcolor}
\usepackage{multirow}
\usepackage{graphicx}
\usepackage{natbib,hyperref}
\usepackage{lineno}
\usepackage{textcomp}

\AtBeginShipout{%
  \ifnum\value{page}>1 %
    \typeout{* Additional boxing of page `\thepage'}%
    \setbox\AtBeginShipoutBox=\hbox{\copy\AtBeginShipoutBox}%
  \fi
}

\definecolor{darkblue}{rgb}{0.0, 0.0, 0.55}
\definecolor{darkred}{HTML}{660022}

\def\apj{ApJ}

\DeclareMathOperator{\asinh}{asinh}

\newcommand{\zphoto}{z_\mathrm{photo}}
\newcommand{\zspec}{z_\mathrm{spec}}
\newcommand{\znorm}{z_\mathrm{norm}}
\newcommand{\fig}[1]{Fig.~\ref{#1}}	
\newcommand{\figFull}[1]{Figure~\ref{#1}}	
\newcommand{\tab}[1]{Tab.~\ref{#1}}

\newcommand{\vect}[1]{\boldsymbol{#1}}

\begin{document}
%
%%%%%%%%%%%%%%%%%%%%%%%%%%%%%%%%%%%%%%%%%%%%%%%%%%%%%%%%%%%%%%%%%%%%%%%%%%%%%%%%

\title{On the realistic validation of photometric redshifts, or why Teddy will never be Happy}

%% Some title propositions:
% BEARPZ: Biased Estimation Analysis of Realistic Photometric Redshifts
% One faint spectrum in hand is worth two bright ones on the sky
% Houston, we need more colors
% Improving photometric redshift estimation with color diversity
% "I see your true colors shining through" (https://www.youtube.com/watch?v=YmsFnSnqoVE)
% Faint spectra matter
% Don't trust your photo-z because your validation set is fooling you
% Impacts of the color-space coverage of the training set on photometric redshift estimation using machine learning
% Realistic validation of photometric redshifts, or why Teddy cannot cover the errors of Happy 
% 

%%%%%%%%%%%%%%%%%%%%%%%%%%%%%%%%%%%%%%%%%%%%%%%%%%%%%%%%%%%%%%%%%%%%%%%%%%%%%%%%

\pagestyle{myheadings}
\markboth{R.~Beck et al.}{Realistic validation of photometric redshifts}

\author[R.~Beck et al.]
{%\parbox{\textwidth}{\vspace{-.5cm} \Large 
%Temporary author list
R.~Beck$^{1}$\thanks{beckrob23@caesar.elte.hu},
C.-A.~Lin$^{2,3}$,
E.~E.~O.~Ishida$^{4}$,
F.~Gieseke$^{5}$,
R.~S.~de Souza$^{6,7}$,
\newauthor
M.~V.~Costa-Duarte$^{7,8}$,
M.~W.~Hattab$^{9}$,
A.~Krone-Martins$^{10}$,
%}
\newauthor for the COIN Collaboration\\
\vspace{0.3cm}\\
\parbox{\textwidth}{ \small 
$^{1}$Department of Physics of Complex Systems, E\"otv\"os Lor\'and University, Budapest 1117, Hungary \\
$^{2}$Service d\textquotesingle Astrophysique, CEA Saclay, Orme des Merisiers, B\^at 709, 91191 Gif-sur-Yvette, France\\
$^{3}$Fenglin Veteran Hospital, 2 Zhongzheng Rd. Sec. 1, Fenglin Township, Hualien 97544, Taiwan\\
$^{4}$Laboratoire de Physique Corpusculaire, Universit\'e Clermont-Auvergne, 
4 Avenue Blaise Pascal, 63178, Aubi\`ere Cedex, France \\
$^{5}$University of Copenhagen, Sigurdsgade 41, 2200 Copenhagen, Denmark\\
$^{6}$MTA E\"otv\"os University, EIRSA ``Lendulet'' Astrophysics Research Group, Budapest 1117, Hungary\\
$^{7}$Instituto de Astronomia, Geof\'isica e Ci\^encias Atmosf\'ericas, Universidade de S\~ao Paulo, R. do Mat\~ao 1226, 05508-090, SP, Brazil\\
$^{8}$Leiden Observatory, Leiden University, Niels Bohrweg 2, 2333 CA Leiden, The Netherlands. \\
$^{9}$Center for Biomarker Research and Personalized Medicine, Virginia Commonwealth University, Richmond, VA, USA\\
$^{10}$CENTRA/SIM, Faculdade de Ci\^encias, Universidade de Lisboa, Ed. C8, Campo Grande, 1749-016, Lisboa, Portugal\\
}
}
\maketitle
\label{firstpage}

%%%%%%%%%%%%%%%%%%%%%%%%%%%%%%%%%%%%%%%%%%%%%%%

\topmargin -1.3cm
% NOTE THAT THIS WILL PLACE THE LAYOUT CORRECTLY ON ASTRO-PH
% PLEASE LEAVE IN

%%%%%%%%%%%%%%%%%%%%%%%%%%%%%%%%%%%%%%%%%%%%%%%%

\begin{abstract} 
Two of the main problems encountered in the development and accurate validation of photometric redshift (photo-\emph{z}) techniques are the lack of spectroscopic coverage in feature space (e.g. colours and magnitudes) and the mismatch between  photometric error distributions associated with the spectroscopic and photometric samples. Although these issues are well known, there is currently no standard benchmark allowing a quantitative analysis of their impact on the final photo-\emph{z} estimation.
 In this work, we present two galaxy catalogues, \textsc{Teddy} and \textsc{Happy}, built to enable a more demanding and realistic test of photo-\emph{z} methods. Using photometry from the \textit{Sloan Digital Sky Survey} and spectroscopy from a collection of sources, we constructed datasets which mimic the biases between the underlying probability distribution of the real spectroscopic and photometric sample.
We demonstrate the potential of these catalogues by submitting them to the scrutiny of different photo-\emph{z} methods, including machine learning (ML) and template fitting approaches. Beyond the expected bad results from most ML algorithms for cases with missing coverage in feature space, we were able to recognize the superiority of global models in the same situation and the general failure across all types of methods when incomplete coverage is convoluted with the presence of photometric errors - a data situation which photo-\emph{z} methods were not trained to deal with up to now and which must be addressed by future large scale surveys. Our catalogues represent the first controlled environment allowing a straightforward implementation of such tests. 
The data are publicly available within the COINtoolbox (\href{https://github.com/COINtoolbox/photoz_catalogues}{https://github.com/COINtoolbox/photoz\_catalogues}).
\end{abstract}


%%%%%%%%%%%%%%%%%%%%%%%%%%%%%%%%%%%%%%%%%%%%%%%%%%%%%%%%%%%%%%%%%%%%%%%%%%%%%%%%
\input{sections/SEC1_introduction}
\input{sections/SEC2_catalogues}

\input{sections/SEC3_photoz}
\input{sections/SEC4_results}
\input{sections/SEC5_discussion}

%%%%%%%%%%%%%%%%%%%%%%%%%%%%%%%%%%%%%%%%%%%%%%%%%%%%%%%%%%%%%%%%%%%%%%%%%%%%%%%%

\section*{Acknowledgements}

This work is a product of the 3$^{\rm rd}$ COIN Residence Program. We thank Zsolt Frei for encouraging the realization of this edition. The program was held in Budapest, Hungary in August/2016 and supported by the Hungarian Academy of Sciences and Astrophysics group at E\"otvos Lorand University. 

We thank Laszlo Dobos, Madhura Killedar, Maria Luiza Linhares Dantas, Nikolay Kolesnikov and Pierre-Yves Lablanche for enlightening discussions during the program.   EEOI thanks Renuka Pampana and Ricardo Vilalta for valuable discussions on domain adaptation. We also thank T. Pl\"ussmaci for helping us establish an unorthodox working environment.

RSS thanks FAPESP (projects 2016/13470-3, 2012/00800-4) for financial support. MVCD thanks his scholarship from FAPESP (processes 2014/18632-6 and 2016/05254-9). RB was supported through the New National Excellence Program of the Ministry of Human Capacities, Hungary.

The IAA Cosmostatistics Initiative\footnote{\href{http://cointoolbox.github.io/}{http://cointoolbox.github.io/}} (COIN) is a non-profit organization whose aim is to nourish the synergy between astrophysics, cosmology, statistics and machine learning communities. We thank John Hammersley and the whole  Overleaf team for providing collaborative tools with which this paper was written\footnote{\href{https://www.overleaf.com/org/coin}{https://www.overleaf.com/org/coin}}. This work  made use of the GitHub\footnote{\href{www.github.com}{www.github.com}},  a web-based hosting service, the \texttt{git} version control software, and Slack\footnote{\href{https://slack.com}{https://slack.com}}, a team collaboration platform.  

In memoriam of Joseph M. Hilbe (30th December, 1944 - 12th March, 2017).

%%%%%%%%%%%%%%%%%%%%%%%%%%%%%%%%%%%%%%%%%%%%%%%%%%%%%%%%%%%%%%%%%%%%%%%%%%%%%%%%
\bibliographystyle{mnras}
\bibliography{ref}

%%%%%%%%%%%%%%%%%%%%%%%%%%%%%%%%%%%%%%%%%%%%%%%%%%%%%%%%%%%%%%%%%%%%%%%%%%%%%%%%
\end{document}

%% file: sections/SEC1_introduction.tex
\section{Introduction}
\label{sec:intro}
%%%%%%%%%%%%%%%%%%%%%%%%%%%%%%%%%%%%%%%%%%%%%%%%%%%%%%%%%%%%%%%%%%%%%%%%%%%%%%%%

Photometric redshift (photo-\emph{z}) estimation has become a widespread and vital tool in the astronomical field. Although compared to their higher resolution counterpart, the spectroscopic technique (spec-\emph{z}), photo-\emph{z} measurements are subject to higher uncertainty, they are also more efficient,  cheaper, and able to probe more distant objects \citep[e.g.][]{Hildebrandt08}. These characteristics make them more suitable for some astrophysical problems. An example of the former is the weak gravitational lensing \citep{2008MNRAS.387..969A}, which measures the coherent galaxy shape distortion by gravitational potentials, and is relatively less sensitive to the redshift measurement. The lensing signal is a convolution of the density contrast distribution with a broad kernel that has the effect to smooth the sensitivity of the signal to the redshift accuracy. Therefore, in order to obtain a faster measurement and to maximize the data volume, photo-\emph{z} is commonly use in weak lensing studies.

However, with the arrival of the Stage-IV lensing surveys \citep{Natarajan2014}, the goal on cosmological constraints becomes more ambitious, and consequently, the requirements on photo-\emph{z} precision and accuracy equally increase. For instance, for the Euclid\footnote{\href{http://sci.esa.int/euclid/}{http://sci.esa.int/euclid/}} mission of the European Space Agency (ESA), the initial requirements on the bias and the scatter in each redshift bin were 0.002 and 0.05 for a total number of 2 billion galaxies, respectively~\citep{Laureijs_etal_2011}. Spectroscopic follow-up  for such a large number of objects is infeasible and such stringent requirements on redshift measurements are extremely challenging for current photo-\emph{z} methods. Apart from weak lensing, other applications exist such as large scale structure \citep{Malavasi2016} and gravitational waves \citep{Antolini2016}, which also require improvements in the current techniques.

In the quest for a viable photo-\emph{z} alternative capable to handle the size and complexity of modern astronomical surveys, a plethora of different methods have been proposed and tested. These are commonly divided into two main classes:
i) template fitting ~\citep[e.g.,][]{benitez2000, bolzonella2000, Csabai2000, ilbert2006, Coe2006, brammer2008, Leistedt2016, beck2016B} ii) empirical \citep[e.g.][]{wadadekar2005, Boris2007, miles2007, Budavari2009, Carliles2010, omill2011, Krone-Martins14a, Elliott2015, Hogan2015, cavuoti2015} and iii) hybrid techniques ~\citep[e.g.][]{Beck2016}.

In template fitting techniques, a set of synthetic spectra is determined from synthesised stellar population models for a given set of metallicities, star formation histories and initial mass functions, among other properties. The photo-\emph{z} is calculated by determining the synthetic photometry (and thus spectral template and redshift) which best fits the photometric observations. Empirical techniques, on the other hand, usually require a data set with spectroscopically measured redshifts in order to train an algorithm which will subsequently be applied to a pure photometric sample. Hybrid methods represent a combination of the previous ones, using an empirical step to first determine the photometric redshift and a template fitting step where physical information provided by the templates can be used to evaluate the accuracy of the photo-\emph{z} determination.

%%%%%%%%%%%%%%%%%%%%%%%%%%%%%%%%%%%%%%%%%%%%%%%%%%%%%%%%%%%%%%%%%%%%%%%%%%%%%
%%%%    Figure  - Colour-magnitude distribution for Teddy
\begin{figure*}
	\includegraphics[width=\textwidth, trim={0 0.2cm 0 0}]{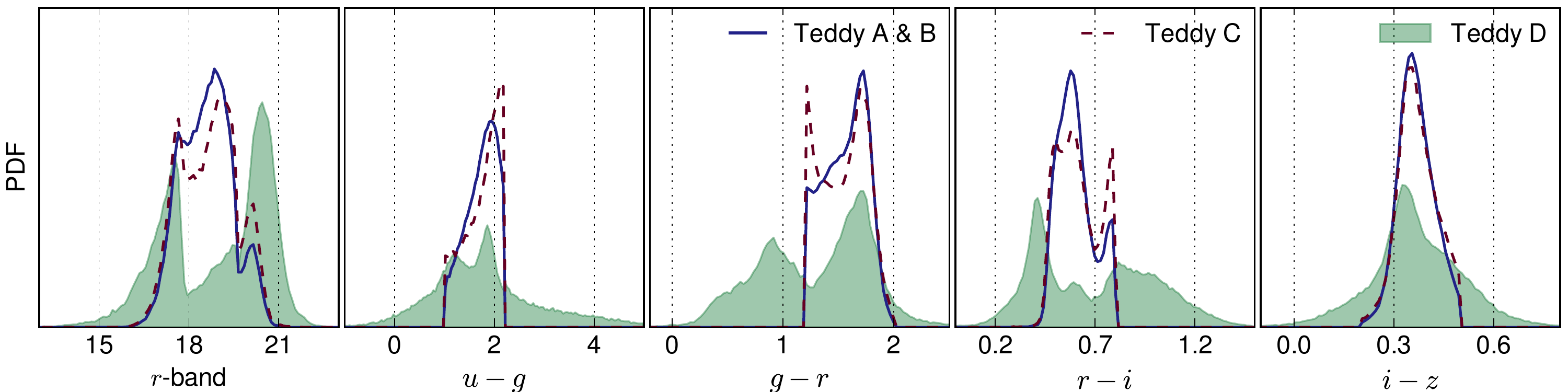}
	\caption{Distributions for $r$-band magnitude and 4 SDSS colours for the \textsc{Teddy} catalogue.}
    \label{fig:colorDistrib_teddy}
\end{figure*}
%%%%%%%%%%%%%%%%%%%%%%%%%%%%%%%%%%%%%%%%%%%%%%%%%%%%%%%%%%%%%%%%%%%%%%%%%%%%%

There have been several notable publications that contrasted the performance of empirical and/or spectral model fitting codes \citep[e.g.,][]{Csabai2003, Hildebrandt2010, Abdalla2011, Dahlen2013}. However, when photo-\emph{z} algorithms are evaluated in the literature, the available spectroscopic set is randomly split into training and validation sets. Roughly, the algorithm is optimized using the training set and its accuracy estimated based on its performance on the validation set. This approach neglects to take into account the fact that the distribution of galaxies in the space of observables (i.e photometric magnitudes/colours) is generally very different for the spectroscopic and the photometric samples, and even their range in the magnitude/colour space can differ. Moreover, given the quality requirements demanded for spectroscopic observation, the photometric sample encloses a larger photometric uncertainty which may break the direct relation between magnitude/colours and redshift,  even when this relation is clearly defined in the spectroscopic sample. In summary, the performance metrics obtained on the former is not representative of results for the latter. In the worst-case scenario, these issues are ignored altogether, as in the benchmark papers aforementioned. In other cases,  error flags or feature selection results are provided along side the photo-\emph{z} estimation, notifying users when extrapolation is performed and results should not be trusted \citep{Brescia2014, Beck2016, stensbo2017} - essentially cutting the coverage of the photometric sample.

Intermediate approaches, aiming at dealing with at least one of these problems have also been reported. In situations where spectroscopic and photometric samples share the same coverage in magnitude/colour space, it is possible to adapt the spectroscopic sample\footnote{In machine learning jargon such methods are a subclass of \textit{Domain Adaptation} techniques.} distribution in order to get it closer to the photometric one \citep[e.g.][]{Lima2008, sanchez2014, Kremer2015}. However, in real-data scenarios, especially when upcoming surveys are considered, even that assumption will not hold. To obtain a measure of photo-\emph{z} performance that is realistic for the actual use case, i.e. when the photometric sample has a much wider coverage in colour space than the spectroscopic sample and, at the same time, there is a correlation between colours and photometric errors, it is crucial that we evaluate photo-\emph{z} methods in more realistic data situations.

In this paper, we provide for the first time a complete benchmark template to allow a  realistic evaluation of the performance of photometric redshift estimators. Using photometry from the \textit{Sloan Digital Sky Survey} Data Release 12 \citep[SDSS-DR12,][]{Alam2015} and spectroscopic redshift measurements from a variety of different sources, we were able to construct validation samples which follow the colour  coverage and shape distribution from the original SDSS-DR12 photometric sample. These enable an unprecedented realistic view into the accuracy of current photo-\emph{z} methods and provide a starting point to the development of new techniques which take these issues into account.

The outline of this paper is as follow. In \S \ref{sec:design} we show how we built our benchmark samples from a combination of spectroscopic and photometric data. In \S \ref{sec:methods} we describe the methods we have used to access the impact of non-representativeness. In \S \ref{sec:results} we compare the summary statistics and performance of different photo-\emph{z} estimators. We present a discussion of our results in \S \ref{sec:disc}.

Throughout the paper, we use SDSS modelMag broad-band magnitudes in the SDSS $\asinh$ magnitude system \citep{Lupton1999}, which have been corrected for Galactic extinction according to \citet{Schlegel1998}.

%% file: sections/SEC2_catalogues.tex
\section{The catalogues}
\label{sec:design}

To enable realistic performance estimation for photo-\emph{z}  methods we present two data sets built to mimic the main causes of non-representativeness between spectroscopic (training) and photometric (test) samples: the disparity in colour-space coverage (\textsc{Teddy}) and the differences between photometric error distributions  (\textsc{Happy}). 
Each catalogue is composed of 4 samples with known spec-\emph{z}: one following the characteristics of the real spectroscopic sample, which should be used for training/calibration purposes  (A) and three holding increasing degrees of non-representativeness of A,  which should be used as  test (or validation) sets (B, C and D).  In what follows, we describe how the catalogues were constructed and the main effects they allow us to probe. 

\subsection{\textsc{Teddy}: the effect of colour coverage}
\label{sec:design:teddy}

%%%%%%%%%%%%%%%%%%%%%%%%%%%%%%%%%%%%%%%%%%%%%%%%%%%%%%%%%%%%%%%%%%%%%%%%%%%%%%%%%%%%%%
%%%%  Table: number of objects in each sample
\begin{table} 
	\centering
	\begin{tabular}{rcccc}
 		& Sample A & Sample B & Sample C & Sample D\\
		\hline  \noalign{\vskip 0.1cm}
		\textsc{Teddy} & 74309 & 74557 & 97980 & 75924\\
		\textsc{Happy} & 74950 & 74900 & 60315 & 74642
	\end{tabular}
	\caption{Numbers of galaxies contained in different samples of the \textsc{Teddy} and \textsc{Happy} catalogues. The total number of galaxies in the SDSS-DR12 spectroscopic sample is $2,040,465$, which we extended to $2,209,299$ (see Sec.~\ref{sec:design:happy}).}
	\label{tab:number_samples}
\end{table}
%%%%%%%%%%%%%%%%%%%%%%%%%%%%%%%%%%%%%%%%%%%%%%%%%%%%%%%%%%%%%%%%%%%%%%%%%%%%%%%%%%%%%%

%%%%%%%%%%%%%%%%%%%%%%%%%%%%%%%%%%%%%%%%%%%%%%%%%%%%%%%%%%%%%%%%%%%%%%%%%%%%%%%%%%%%%%
%%%% Figure: magnitude and colour distribution for Happy
\begin{figure*}
\centering
\includegraphics[width=\textwidth]{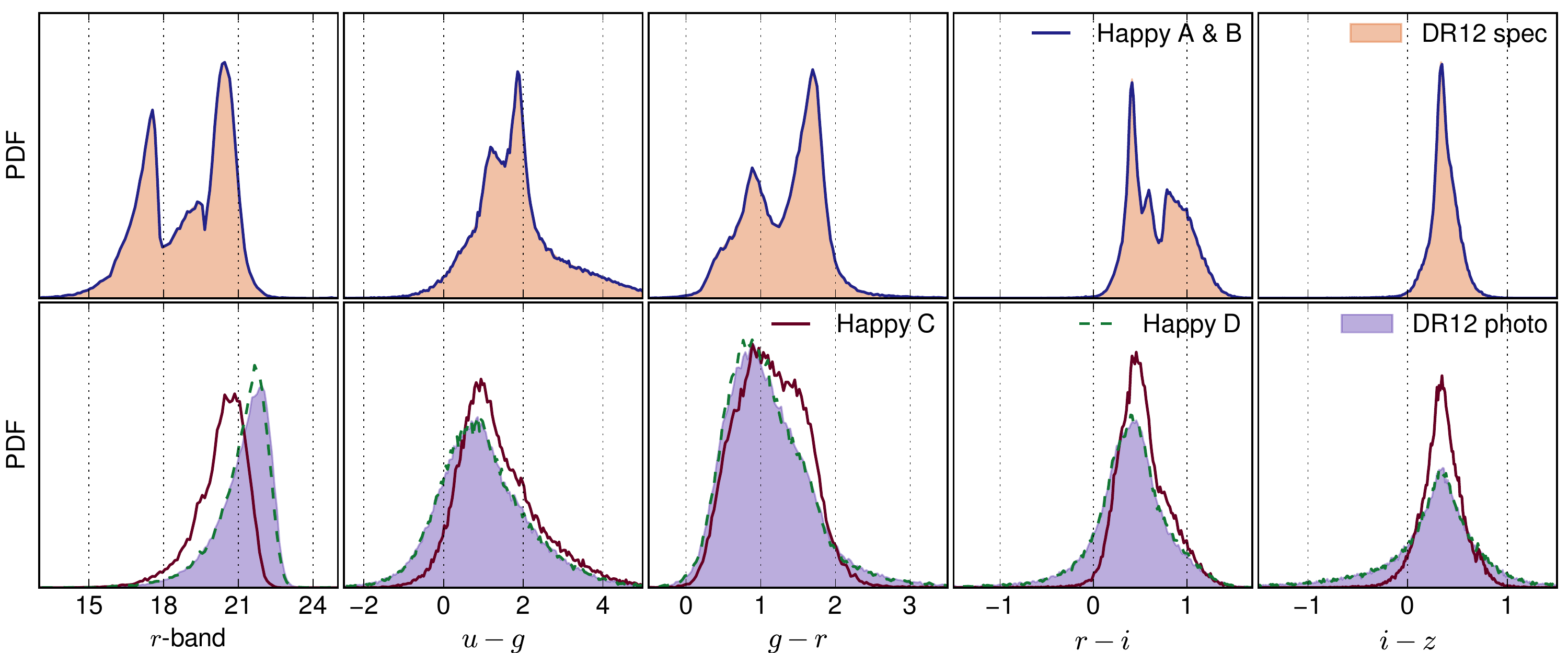}
\caption{Distributions for magnitude in $r$-band and 4 colors for the \textsc{Happy} catalogue compared with the full spectroscopic (red) and photometric (purple) distributions from SDSS-DR12.}
\label{fig:colorDistrib_happy}
\end{figure*}
%%%%%%%%%%%%%%%%%%%%%%%%%%%%%%%%%%%%%%%%%%%%%%%%%%%%%%%%%%%%%%%%%%%%%%%%%%%%%%%%%%%%%%

The disparity in feature-space distribution between training and test sets has been known to impact classification and regression tasks in machine learning \citep{Quionero2009}. Specifically in the photo-\emph{z} case, this translates into a spectroscopic sample which fails to cover the entire domain occupied by the photometric sample in colour-magnitude parameter space. It has been already reported that such a gap introduces significant biases in redshift estimation for empirical \citep[e.g.][]{MacDonald2010} as well as for template fitting techniques \citep[e.g.][]{Dahlen2013}. 

In the context of empirical methods, in which there is no input information beyond magnitudes and colours from the spectroscopic sample, it is straightforward to expect that results will not be reliable beyond the training domain. Machine learning methods only learn by example, thus their results should not be extrapolated. For template fitting methods, the region of the parameter space not covered by the spectroscopic sample can be directly related to fainter objects having active galactic nucleus (AGN) activity and metallicity levels not represented in the template library \citep{MacDonald2010}. Although these issues are widely known, at the moment there is no standard ground to quantify the sensitivity of each photo-\emph{z} method in the presence of such gap.  The \textsc{Teddy} catalogue presented here was designed to fulfil this role. It was completely built using SDSS-DR12 spectroscopic sample, which ensures the quality of photometric measurements and allow us to isolate the effect of colour-magnitude coverage from its correlation with photometric errors. 

For each of the four SDSS colours, we defined intervals forming a 4-dimensional rectangular parallelepiped ($1.0 < u-g < 2.2$, $1.2 < g-r < 2.0$, $0.1 < r-i < 0.8$, and $0.2 < i-z < 0.5$). We then selected $\approx 150,000$ objects from the SDSS-DR12 spectroscopic sample whose colours were within this space and with an extra constraint on $r$-band magnitude, $r < 21$ - this is called the \textit{narrow set}. All other objects were grouped into what we called the \textit{wide set}.

From the narrow and wide sets, we constructed one training sample (named A) and three test samples (named B, C, and D) for comparing different scenarios. Sample A and B were created by splitting the narrow set into two comparable parts, sample C was constructed by truncating the wide set on the colour coverage of the narrow set and finally, sample D was made by sampling randomly from the wide set. The number of objects contained in each sample is shown in \tab{tab:number_samples}. In summary, sample A and B follow exactly the same feature space distribution, sample C has the same coverage of A, but a different distribution, and sample D has a wider coverage in colour-magnitude space. 

The $r$-band magnitude and colour distributions for all samples in the \textsc{Teddy} catalogue are shown in  \fig{fig:colorDistrib_teddy}. We observe at ease the colour cut imposed on samples A, B, and C. By construction, samples C and D follow the same distribution in the space region, but their probability density functions (PDFs) differ due to marginalization. A significant number of galaxies with colours outside the 4-dimensional parallelepiped contribute to the disparity between the two curves.

We consider A the spectroscopic sample (used for training) and  B, C and D as  distinct photometric samples (used for validation/test). 
These correspond to increasingly complex data situations. Training in A and testing in B represents the best case scenario where distributions are completely representative of each other - and thus must yield the best possible results. Training in A and testing in C corresponds to the situation of simply ignoring data outside spectroscopic  coverage - in this case methods like applying weights should provide slightly better results than the standard approach. Training in A and testing in D corresponds to a situation with incomplete coverage, where pure machine learning methods are not expected to provide good results.

It is important to emphasize that the colour and magnitude distributions in \textsc{Teddy} are not realistic, but they do provide a test bench to probe the robustness of photo-\emph{z} methods against feature distribution shape and coverage. For an ideal photo-\emph{z} estimator, results from all three configurations should be equivalent.

\subsection{\textsc{Happy}: the effect of photometric errors}
\label{sec:design:happy}

Once we have a data set which enables probing the robustness of photo-\emph{z} methods in the case of incomplete colour and $r$-magnitude coverage, we approach other important differences  between spectroscopic and photometric samples: the presence of  photometric errors and their correlation with colour coverage. 
In constructing \textsc{Teddy}, we rearranged data from the original SDSS-DR12 spectroscopic data set. Thus, all its samples share the same spectroscopic data quality level. This is not what happens in the real situation, where photometric observations are statistically fainter and of poorer quality than the spectroscopic ones.

The \textsc{Happy} catalogue was created to allow a clear assessment of the impact of photometric errors on photo-\emph{z} estimation, while at the same time more closely resembling the colour space differences between the original SDSS-DR12 spectroscopic and photometric data sets.

Our first goal was to reproduce the colour-magnitude space distribution of the SDSS-DR12 photometric 
sample, only with objects with measured spectroscopic redshifts, so that the photo-\emph{z} methods could be properly evaluated. However, as the DR12 spectroscopic set does not contain objects with more extreme colours and fainter magnitudes, it is not an adequate source of example galaxies. Thus, we chose to extend the DR12 spectroscopic set (S1) by cross-matching SDSS photometric measurements with galaxies from other spectroscopic surveys. This approach provides a deeper sample of spectroscopic galaxies, while also keeping the use of actual SDSS photometry and its inherent systematics.

We followed the Bayesian cross-matching methodology described in \citet{Budavari2008}, only accepting relatively secure matches, with a Bayes factor above $10,000$ \citep[Eq. 16 in][]{Budavari2008}. The following surveys were used in the match:
\begin{itemize}
	\item 2dF \citep[][770 matches]{Colless2001, Colless2003}, 
    \item 6dF \citep[][765 matches]{Jones2004, Jones2009}, 
    \item DEEP2 \citep[][7456 matches]{Davis2003, Newman2013}, 
    \item GAMA \citep[][53373 matches]{Driver2011, Baldry2014}, 
    \item PRIMUS \citep[][32459 matches]{Coil2011, Cool2013}, 
    \item VIPERS \citep[][18967 matches]{Garilli2014, Guzzo2014}, 
    \item VVDS \citep[][8381 matches]{LeFevre2004, Garilli2008}, 
    \item WiggleZ \citep[][43874 matches]{Drinkwater2010, Parkinson2012}, and 
    \item zCOSMOS \citep[][2789 matches]{Lilly2007, Lilly2009}. 
\end{itemize}

\noindent Refer to \citet{Beck2016} for additional details regarding the cross-match -- the same data were used here, with the important distinction that photometric colour and error cuts were not applied.

This procedure enabled us to find $168,834$ matches, which extended the total number of galaxies with spectroscopic redshifts to $2,209,299$, and also extended the colour-magnitude space coverage of the sample such that the parameter range of the SDSS photometric set was covered.

In order to build, from the new extended spectroscopic sample (E1),  a subset which follows the same colour--magnitude distribution as the original SDSS-DR12 photometric set, we randomly selected $75,000$ objects from the SDSS-DR12 photometric sample (S2)\footnote{To be precise, the objects were selected from a $2$ million element random sub-sample of S2.} and performed a nearest neighbour (NN) search in E1 (in colour/r-magnitude space).

For each object in S2 we search for its 1\textsuperscript{st} NN to include into our new set, \textsc{Happy D}. To avoid duplicate entries, if the given NN was already included, we select the next closest NN (2\textsuperscript{nd}, 3\textsuperscript{rd}, etc.) which was not already in \textsc{Happy D}.

Then, we similarly constructed two new subsets to represent the DR12 spectroscopic sample, \textsc{Happy A} and \textsc{Happy B}. The former will act as a training set/spectroscopic sample, while the latter will be a test set that has the same distribution of photometric properties as the training set. Thus, we randomly selected $2 \times 75,000$ objects from S1, and searched for their nearest neighbour in E1 using the method outline above, again avoiding any duplicates within \textsc{Happy} sets.

Finally, to create an intermediate sample that is between the photometric error properties of S1 and S2, we decided to perform a photometric error cut. Our goal was to reproduce the same range of photometric errors as in S1, but with a distribution that resembles S2, being more weighted towards higher errors. Thus, the cut was chosen to be at the 98\textsuperscript{th} percentile (to discard outliers) of the photometric error distribution of S1 for each observed feature. We randomly selected $150,000$ objects from S2, searched for their nearest neighbours in E1 following the same procedure, and applied the error cut. This yielded the set \textsc{Happy C}, composed of $\approx 60,000$ galaxies. We note that contrary to all other \textsc{Happy} set pairings, \textsc{Happy C} and \textsc{Happy D} were allowed to overlap to avoid excessively selecting from the less populated faint end of E1.

Cutting the photometric error range of \textsc{Happy C} to match that of \textsc{Happy A} and \textsc{B} had an important side effect: the magnitude and colour ranges (note: the range, not the shape of the distribution) were also essentially cut to the limits of \textsc{Happy A} and \textsc{B}. This shows that the effects of photometric error and magnitude/colour coverage are in fact very much intertwined, one cannot modify one without affecting the other. There are two feasible explanations for why the colour ranges shrink because of the error cut. First, this observation could indicate that the wider colour distributions in the photometric sample are mainly caused by the higher photometric errors smearing the distribution, not by containing physically different extreme galaxies that are missing from the spectroscopic sample. Second, it could be a consequence of galaxy types with extreme colours being significantly more likely to have high measurement errors, therefore these would be preferentially eliminated by the cut, and also wouldn't be present in the spectroscopic sample.
Figure \ref{fig:colorDistrib_happy} shows the $r$-magnitude and colour PDFs for all samples in \textsc{Happy}, and the number of objects contained in each set is shown in \tab{tab:number_samples}.

Following the same reasoning as presented in the previous section, we built \textsc{Happy A} to act as a spectroscopic sample and, as such, should be used for training. \textsc{Happy B} is completely representative of \textsc{Happy A} and so mirrors the equivalent scenario of traditional photo-\emph{z} validation exercises. 
\textsc{Happy C} illustrates a photometric sample that has been cut to conform to the training sample, with a larger proportion of objects having high photometric errors. \textsc{Happy D} serves as a complete photometric sample, with both a wider parameter coverage and higher measurements errors. Thus, \textsc{Happy C} and \textsc{D} must only be used for testing, representing increasing degrees of complexity and similarity with the real photometric situation.

%% file: sections/SEC3_photoz.tex
\section{Photometric redshift estimation}
\label{sec:methods}

Aiming at quantifying the impact of the above discussed effects in traditional photo-\emph{z} estimation methods, we selected a few examples to illustrate how the colour--magnitude coverage and its correlations with photometric redshift accuracy can be quantified in future discussions on photo-\emph{z} methods.

\input{sections/SUBSEC_3.1_empirical}

\input{sections/SUBSEC_3.2_template}

%% file: sections/SUBSEC_3.1_empirical.tex
\subsection{Empirical methods}

\label{sec:methods_empirical}

In what follows, we introduce a selection of empirical photo-\emph{z} estimation techniques that were chosen to represent this class of methods.

\subsubsection{Artificial neural network (ANNz)}

\textsc{ANNz} \citep{Collister_Lahav_2004} implements a particular type of artificial neural networks known as multi-layer perceptron, which is formed by a set of layers, each one of them populated by a number of nodes. The first layer receives the observed magnitudes, or colours, the final layer returns the estimated photo-\emph{z} values and the intermediate layers are considered hidden - since they can contain any number of nodes. All nodes in a given layer hold an activation function and are connected to all the nodes in adjacent layers, with each connection corresponding to a weight parameter $w_{i,j}$\footnote{The indexes indicate the two nodes connected by this parameter.}. 

Given a training set with measured magnitudes and spectroscopic redshifts, the network is trained by determining the set weight parameters values, $\pmb{w}$, which minimizes the cost function
\begin{equation}
E=\sum_{i=1}^N\left[z_{p}(\pmb{w}, \bar{m}) - z_{spec}\right]^2,
\end{equation}
where $\bar{m}$ is the set of observed magnitudes, $z_p$ is the output given by the last layer and $z_{spec}$ the spectroscopic redshift \citep{Abdalla2011}. 

The neural network used in this study is configured to have two intermediate layers, resulting in four layers in total. The first layer receives five input features: the $r$-band magnitude and four colors, normalized by their respective mean and standard deviation. The two intermediate layers contain each ten nodes. The final layer outputs one node for the photo-\emph{z} prediction, so that the whole network has a 5-10-10-1 structure.

In what follows, the network was trained in \textsc{Teddy A} and \textsc{Happy A} and subsequently applied to the other samples in each catalogue. We did not consider measurement errors in this work.

\subsubsection{Local linear regression (LLR)}

The local linear regression method finds the $k$-nearest neighbours of the test galaxy in colour space from a training set, and performs a hyperplane fit on these neighbours. This way, a functional form is fitted that also follows the local trends, and therefore can be quite flexible. The implementation used here is the same as in \citet{Beck2016}, refer to that paper for more details. We note that while \citet{Beck2016} performed a template fitting step after the photo-\emph{z} estimation itself, in this paper we do not utilize the extra physical information, therefore that step was omitted entirely. Thus, the method used here can be categorized as purely empirical.

The number of nearest neighbours used in \citet{Beck2016} is $k=100$, but since we are also testing extrapolation capabilities, that number was increased to $k=1000$ here. The increase does not noticeably affect estimation results when within the coverage of the training set, but taking a larger colour space region into account better determines the functional form, and significantly reduces scatter when extrapolating. Of course, there is a trade-off in computational performance when using more neighbours.

\subsubsection{Generalized additive model (GAM)}

Generalized linear models (GLMs), as its name suggests, assume - through a link function - a linear relationship between the response variable $y$ and set of predictors $\vect x$. The distribution of $y$ is a member of the exponential family and the  link function is monotonic and differentiable. However, GLMs also allow different relationships between the mean and the variance. For example, in linear models (LMs) the response mean is independent of the variance and given simply by $E(y) = \vect x^T \vect \beta$. For Poisson models with log link $ \log E(y)  = \vect x^T \vect \beta $, the mean is equal to the variance. In this context, $\vect \beta$  is a vector of  unobservable regression coefficients to be estimated from the data using Maximum Likelihood (ML) methods. 

Let $\hat \beta$ be the ML estimate of $\vect \beta$. A prediction at a new point $\vect x_0$ is obtained by $\vect x_0^T\hat{\vect \beta}$ for LMs and by $\exp( \vect x_0^T\hat{\vect \beta})$ for Poisson models. GLMs also enclose logistic and gamma regression, among many others. For the following discussion we will only consider LMs but the same reasoning can be followed for any member of GLMs. For a comprehensive treatment on LMs see \citet{Isobe1990,kutner2005applied,dobson2010introduction, christensen2011plane, myers2012generalized}  and for GLMs see \citet{nel72,dobson2010introduction,myers2012generalized}. For GLMs applications in astronomy the reader is refereed to   \citet[e.g.][]{and10,deSouza2014,deSouza2015,Elliott2015,deSouza2016}.
  
  It is understood that the exact functional relationship between $E(y)$ and $\vect x$ is unknown. In other words, $E(y) = f(\vect x) $ for unknown $f$. LMs assume that $f$ can be approximated by $\vect x^T\vect \beta$. Despite its simplicity, LMs have been successfully applied in many areas. However, it has been found that the linearity assumption can be restrictive and too simple to account for non-trivial relationships in the data. Non-parametric regression relaxes this assumption and allows to estimate $f$ directly without imposing any specific functional formula. In fact, $f(\vect x) = \sum_{k=1}^{\infty} \alpha_k \phi_k(\vect x) $ where $\phi_k$'s are known basis functions. Restricting the upper bound of $k$ to a reasonable number $K$, $f(\vect x) \approx \sum_{k=1}^{K} \alpha_k \phi_k(\vect x)$. Then one can estimate $\alpha$'s as usual using parametric regression methods.  Examples of basis functions include cubic spline basis, B-splines, Haar wavelet basis functions and radial basis functions. 
    
This set-up does not impose any assumption on $f$. Even with a modest number of predictors, this attractive property requires estimating a huge number of parameters. Thus $f$ cannot be estimated properly due to the curse of dimensionality. To tackle this problem, \citet{hastie1990} suggests to fit generalized additive models (GAMs). 

GAMs assume that $f(\vect x ) = f_1( x_1)  + f_2(x_2) + \hdots +f_p( x_p)$. If each $f_j$ demands $m$ components then the total number of parameters is $p \times m$ which is reasonable even for moderate size studies. Penalized least squares \citep[e.g.][]{ruppert2003semiparametric} is a powerful approach to fit GAMs. Traditionally, GAMs are fitted using backfitting algorithm \citep{hastie1990}. If each $f_j$ estimated by $\hat f_j$ then $f(\vect x)$ can be estimated by $\hat f(\vect x) = \sum_{j=1}^{p} \hat f_j (x_j)$. A prediction at $\vect x_0$ is obtained by $\hat f(\vect x_0) = \sum_{j=1}^{p} \hat f_j (x_{0j})$. Additional details on fitting models and conducting inferential statistics using GAMS can be found in \citep{hastie1990,ruppert2003semiparametric,wood2006generalized}. We should note that the GAM methodology herein developed is a more general extension of the {\sc cosmophotoz} \citep{Elliott2015} package, who first introduced the use of GLMs for redshift estimation.

\subsubsection{Random forest}

Random forests depict ensembles of individual classification or regression trees, which are fit given the available training data. Each tree is built from top to bottom, where the root corresponds to all considered training instances and the leaves to subsets of instances. The internal nodes of each tree are usually split recursively (into two) based on certain splitting criteria. The overall recursive process stops as soon as some stopping criterion is fulfilled per leaf node. In case of fully grown trees, each leaf corresponds to single pattern or to a group of ``pure'' patterns (i.e., a group of patterns having all the same label).

The original way of constructing random forests is to consider, for each individual tree, a subset of the training patterns, called \emph{bootstrap sample}~\citep{Breiman2001}. These bootstrap samples are drawn uniformly at random (with replacement) to generate slightly different training sets and, hence, slightly different individual trees. Another way is to consider slightly different splits at each internal node by, e.g., considering different feature dimensions or random splitting thresholds. The overall performance of such a forest of trees is usually much better than the one of the individual trees (due to the variance reduction that stems from combining the predictions made by the trees).

The splitting processes taking place at the internal nodes is based on measuring the gain in ``purity'', which can be, in turn, measured via different scores. Typical ones, measuring how impure a set of patterns associated with a node is, are the \emph{mean squared error}~(MSE) for regression problems or the \emph{Gini index} for classification problems. We refer to~\cite{Breiman2001} for a detailed description.

In this work, we consider random forest regressors as we are interested in real-valued redshift estimates. While various parameters can be set for random forests, the performances of the induced models are often very similar among each other as long as reasonable parameter assignments are chosen. In the remainder of this work, we consider the following setup: For all random forest models, 500 individual fully-grown trees are fitted given bootstrap samples, where $\sqrt d$ features are tested per internal node split using the MSE as impurity measure.

%n_estimators = 500
%n_jobs = 2
%min_samples_split = 2
%criterion = "mse"
%max_features = "sqrt"
%bootstrap = True

%% file: sections/SUBSEC_3.2_template.tex
\subsection{Template fitting methods}

\label{sec:methods_template}

In this section, we continue on to outlining the details of the spectral template fitting methods  that were analysed in this paper.

\subsubsection{Bayesian Photometric Redshifts (BPZ)}

\label{sec:methods_template_BPZ}

BPZ\footnote{\href{http://www.stsci.edu/~dcoe/BPZ/}{http://www.stsci.edu/$\sim$dcoe/BPZ/}} applies Bayesian inference to the problem of photometric redshift estimation \citep{benitez2000}. In this context, the probability of a galaxy with measured colour and magnitudes, $\{C, \pmb{m}\}$, to have a redshift $z$ can be described as
\begin{equation}
p(z|C, \pmb{m})\propto p(z|\pmb{m})p(C|z),
\end{equation}
where $p(C|z)$ is the probability of the observed colours given a galaxy at redshift $z$ (likelihood) and $p(z|\pmb{m})$ is the expected redshift distribution for galaxies with measured magnitudes $\pmb{m}$ (prior). This description assumes that the likelihood depends only on the measured magnitudes and morphological galaxy type \citep{benitez2000}. The feature which differentiates this approach from the others which came before it is the introduction of the prior. It improves over the simple template fitting $\chi^2$ minimization by allowing the introduction of information about the galaxy morphological type and helps avoiding unrealistic redshift estimations by using simple assumptions as the range of redshift expected for a particular survey.

In this work we present results using the default set of 8 spectral energy distributions (SEDs) based on \citet{coleman1980} and \citet{kinney1996} and two extra interpolated ones between each pair (default option). The filters zero-points  were calibrated using samples  \textsc{A}.

\subsubsection{EAZY}

 \emph{Easy and Accurate Redshifts from Yale}\footnote{\href{http://www.astro.yale.edu/eazy/}{http://www.astro.yale.edu/eazy/}} (EAZY) %represents an attempt to minimize the dependence of template  fitting methods on the spectroscopic sample \citep{brammer2008}. 
%It eliminates the calibration issues 
minimizes the dependence on spectroscopically measured spectra by relying on synthetic SEDs from semi-analytical models. This set, despite not enclosing all possible galaxy types and dust models, provides further completeness to UV and NIR wavelengths than SEDs built from spectroscopic observations \citep{brammer2008}.
 
The default implementation, used in this work, constructs a minimum representative template set of 5  spectra derived from the application of a \emph{non-negative matrix factorization} (NMF - \citet{blaton2007}) algorithm to the set of 485 synthetic spectra provided by \citet{bruzual2003}.
NMF can be considered a kind of ``principal component''  derivation, with the additional constraint that the linear combination coefficients need to be non-negative. Results are thus more interpretable than the ones derived from a standard principal component analysis \citep{Ishida2011A&A,Ishida2013MNRAS,jolliffe2013,2014MNRAS.440..240D}.

%In the context of this paper, EAZY provides an interesting perspective on the role played by the spectroscopic sample in photo-\emph{z}\ determination.
%\rbeck{(COMMENT: I cut a sentence here -- as I understand, they do not use the spectroscopic sample when getting their 5 basis vector spectra, so as long as no calibration is performed for the filters, they aren't using the spec sample at all, just like other template fitting methods.)}
Although the templates used in this method do provide a larger wavelength coverage, it is important to emphasize that the semi-analytical models themselves are constructed from detailed visual and NIR observations of nearby objects. Thus, the accuracy of these models in predicting the spectral behaviour of high redshift galaxies is also limited. As the authors pointed out  themselves, the discrepancy  in UV and NIR  fluxes between  observed spectra and the one chosen by EAZY is larger in the rest-frame UV and NIR wavelengths. Consequently, its results will also be subjected to the lack of representativeness discussed above.

\subsubsection{Photo-z-SQL}

Photo-z-SQL\footnote{\href{https://github.com/beckrob/Photo-z-SQL}{https://github.com/beckrob/Photo-z-SQL}} is a recent Bayesian template fitting photo-\emph{z}  implementation in C\# that can be integrated into a database running Microsoft SQL Server \citep{beck2016B}. The code is fairly flexible in the choice of templates and priors and thus can easily adopt successful approaches found in the literature. Moreover, it features an iterative photometric zero-point calibration to optionally take into account a spectroscopic training set.

We used the stand-alone version of the code, searching for the maximum probability photo-\emph{z}  value using Bayesian estimation. We present results for two configurations, both computed on a redshift grid with a linear step size of $0.01$ between $z=0$ and $z=1$.

The first configuration, which we refer to as SQL BPZ, uses the BPZ spectral template set of \citet{Coe2006}, and the prior of \citet{benitez2000} that had been derived from HDF-N data. It also utilizes an empirical filter zero point calibration based on the training sets, and adds a separate photometric error term of $0.02$ mags to account for template mismatch. The two major differences between this and the earlier BPZ approach (\S\ref{sec:methods_template_BPZ}) are the differing calibration, and the larger number of interpolated templates (ten between each pair).

The second configuration -- denoted by SQL LP -- uses the Le PHARE spectral templates of \citet{Ilbert2009} in conjunction with a flat prior. In this case, zero point calibration was not used, and the extra error term was chosen to be only $0.01$ mags due to the larger and more detailed set of templates (641 for SQL LP, as opposed to 71 for SQL BPZ).

These configurations were selected to optimize results on the \textsc{Teddy B} and \textsc{Happy B} samples, which correspond to the usual case of only having validation information based on a spectroscopic set. In those samples, the SQL LP configuration did not benefit from either the calibration or applying the HDF-N prior, while the SQL BPZ case was improved by both.

For reference, the SQL BPZ and SQL LP configurations correspond to the notation \textit{BPZ HDF Err ZP} and \textit{LP Flat Err}, respectively, in \citet{beck2016B}.

%% file: sections/SEC4_results.tex
\section{Results} 
\label{sec:results}

\input{sections/SUBSEC_4.1_diagnostics}
\input{sections/SUBSEC_4.2_teddy}
\input{sections/SUBSEC_4.3_happy}

\input{sections/SUBSEC_4.4_DA}
%%%%%%%%%%%%%%%%%%%%%%%%%%%%%%%%%%%%%%%%%%%%%%%%%%%%%%%%%%%%%%%%%%%%%%%%%%%%%%%%

%% file: sections/SUBSEC_4.1_diagnostics.tex
\subsection{Diagnostics of estimators}
Following earlier works that compare photo-$z$ methods \citep{Hildebrandt2010,Dahlen2013}, we selected four summary statistics to quantify the photo-$z$ estimation quality of the various methods tested here. We consider the normalized redshift error, $\Delta\znorm=(\zspec-\zphoto)/(1+\zspec)$, and from the distribution of $\Delta\znorm$, we compute its mean (which is also the average bias), standard deviation (std), median absolute deviation (MAD), and outlier rate. Outliers are defined by $|\Delta\znorm|>0.15$. The median absolute deviation $\mathrm{MAD}=\mathrm{median}\left(|\Delta\znorm|\right)$ is computed with outliers included; and the mean $\overline{\Delta\znorm}$ and standard deviation $\sigma\left(\Delta\znorm\right)$ are computed with the outliers removed from the samples. 

%% file: sections/SUBSEC_4.2_teddy.tex
\subsection{Results from Teddy}

In order to quantify the impact of the lack of $r$-magnitude/colour coverage between  spectroscopic and photometric samples, we applied the photo-\emph{z} methods described above to the \textsc{Teddy} catalogue. Methods were trained/calibrated on \textsc{Teddy A} and tested on \textsc{Teddy B, C} and \textsc{D} to represent increasing levels of mismatch.

%%%%%%%%%%%%  Teddy table  %%%%%%%%%%%%
\begin{table*} 
	\centering
    \begin{minipage}{0.49\textwidth}
	\centering
	\begin{tabular}{cccccc}
    	\hline\hline\\[-2.25ex]
		\multirow{3}{*}{Method} & \multirow{3}{*}{Set} & \multicolumn{4}{c}{Diagnostics}\\[0.5ex]
 		&  & Mean & Std & MAD & Outlier rate\\ 
        & & ($\times 10^{-2}$) & ($\times 10^{-2}$) & ($\times 10^{-2}$) & (\%) \\
		\hline \noalign{\vspace{0.1cm}}
		\multirow{3}{*}{ANNz}  & B & 0.03 & 2.35 & 1.16 & 0.18\\ 
 		& C & -0.01 & 2.45 & 1.15 & 0.26\\ 
 		& D & -0.08 & 5.67 & 3.61 & 3.09\\ 
		\hline \noalign{\vspace{0.1cm}} 
		\multirow{3}{*}{GAM}  & B & 0.05 & 2.62 & 1.34 & 0.11\\ 
 		& C & 0.06 & 2.79 & 1.38 & 0.18\\ 
 		& D & -0.06 & 3.93 & 2.23 & 2.28\\ 
		\hline \noalign{\vspace{0.1cm}} 
		\multirow{3}{*}{LLR}  & B & 0.07 & 2.35 & 1.14 & 0.19\\ 
 		& C & 0.05 & 2.44 & 1.14 & 0.28\\ 
 		& D & 1.76 & 4.08 & 2.46 & 3.80\\ 
		\hline \noalign{\vspace{0.1cm}} 
		\multirow{3}{*}{\parbox{1.1cm}{\centering {\footnotesize Random}\\{\footnotesize forest}}}  & B & 0.03 & 2.38 & 1.18 & 0.17\\ 
 		& C & -0.01 & 2.49 & 1.17 & 0.26\\ 
 		& D & 0.16 & 6.85 & 5.24 & 6.70\\
        \hline
	\end{tabular}
	\end{minipage}
    \begin{minipage}{0.49\textwidth}
	\centering
	\begin{tabular}{|cccccc}
    	\hline\hline\\[-2.25ex]
		\multirow{3}{*}{Method} & \multirow{3}{*}{Set} & \multicolumn{4}{c}{Diagnostics}\\[0.5ex]
 		&  & Mean & Std & MAD & Outlier rate\\
        &  & ($\times 10^{-2}$) & ($\times 10^{-2}$) & ($\times 10^{-2}$) & (\%) \\
		\hline \noalign{\vspace{0.1cm}} 
		\multirow{3}{*}{BPZ}  & B & 3.51 & 3.07 & 3.63 & 0.49\\ 
                              & C & 3.48 & 3.30 & 3.69 & 0.58\\ 
                              & D & 2.61 & 4.73 & 3.66 & 3.60\\ 
        \hline \noalign{\vspace{0.1cm}} 
        \multirow{3}{*}{EAZY}  & B & -2.99 & 3.82 & 3.05 & 2.71\\ 
                               & C & -3.71 & 4.07 & 3.57 & 4.03\\ 
                               & D & -3.64 & 4.62 & 4.34 & 6.58\\ 
        \hline \noalign{\vspace{0.1cm}} 
        \multirow{3}{*}{SQL BPZ}  & B & 2.13 & 3.34 & 2.28 & 0.43\\ 
                                  & C & 1.68 & 3.40 & 2.00 & 0.64\\ 
                                  & D & 0.94 & 4.06 & 2.41 & 2.01\\ 
        \hline \noalign{\vspace{0.1cm}} 
        \multirow{3}{*}{SQL LP}  & B & -0.45 & 3.40 & 1.95 & 0.53\\ 
                                      & C & -0.7 & 3.61 & 2.14 & 0.70\\ 
                                      & D & -0.48 & 4.19 & 2.74 & 3.29\\ 
    \hline
	\end{tabular}
	\end{minipage}
	\caption{Results for the \textsc{Teddy} catalogue.}
	\label{tab:Teddy_results}
\end{table*}

%\parbox{1.3cm}{\centering {\footnotesize SQL}\\{\footnotesize BPZ}}
%\parbox{1.3cm}{\centering {\footnotesize SQL}\\{\footnotesize LePhare}}

\subsubsection{Empirical methods}
%%%%%%%%%%%%%%%%%%%%%%%%%%%%%%%%%%%%%%%%%%%%%%%%%%%%%%%%%%%%%%%

\label{sec:teddy_ML}

%%%%%%%%%%%%  Teddy figure  %%%%%%%%%%%%
\begin{figure*}
	\centering
	\includegraphics[width=0.95\textwidth]{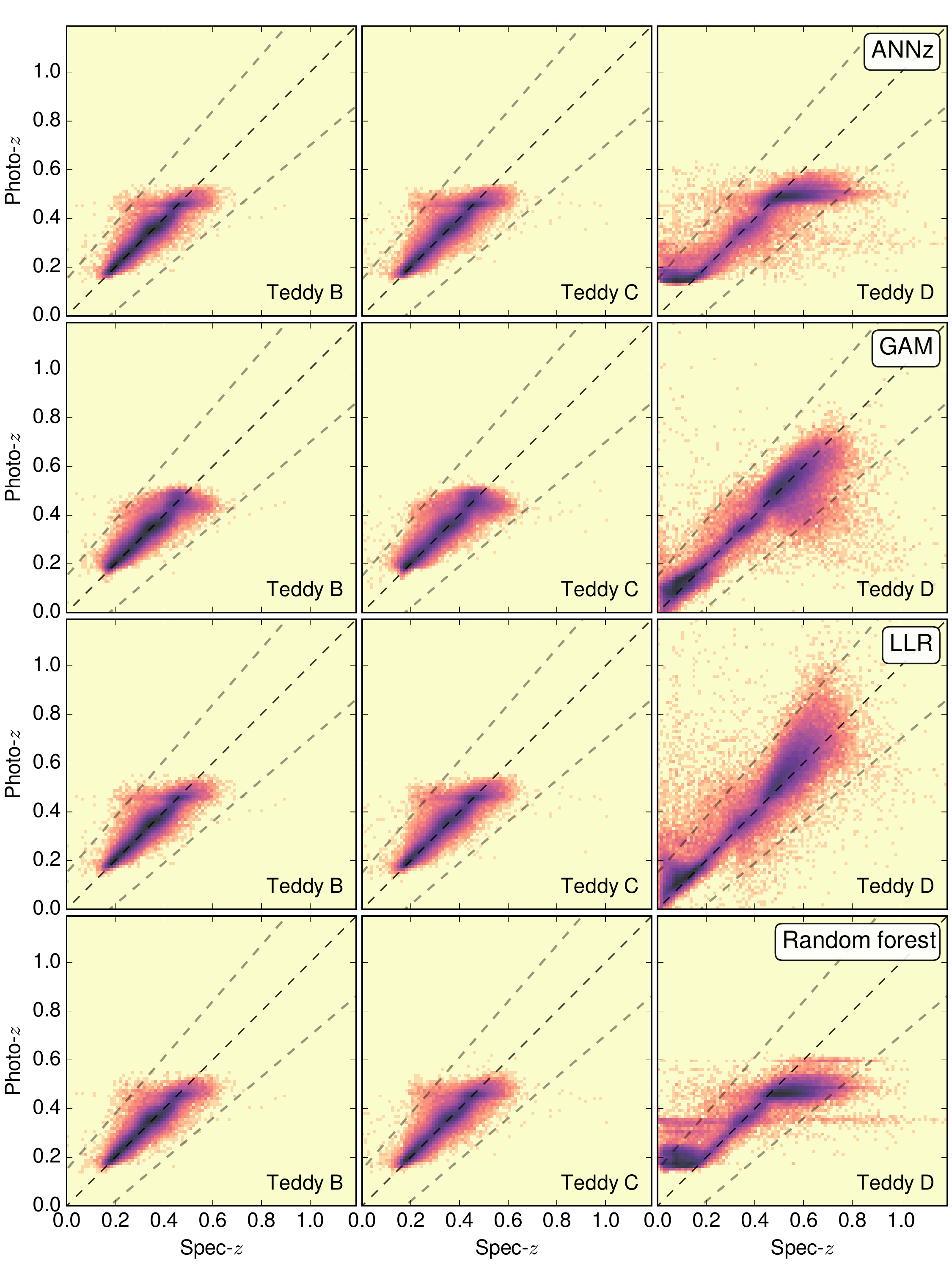}
	\caption{Results on three testing sets of the \textsc{Teddy} catalogue (columns) obtained from four empirical photo-\emph{z} methods (lines). The colour gradient shows the logarithmic density. The dashed lines define the perfect prediction (center) and the limits for being considered outliers. Numerical results are shown in \tab{tab:Teddy_results} - left panel.}
	\label{fig:TeddyRes}
\end{figure*}
%%%%%%%%%%%%%%%%%%%%%%%%%%%%%%%%%%%%%%%%%

% Set B - machine learning
\figFull{fig:TeddyRes} shows the photo-\emph{z} estimations versus their spec-\emph{z} values. Each row represents one of the four machine-learning methods described in \S\ref{sec:methods_empirical}. \textsc{Teddy B}, represented on the left panels of \fig{fig:TeddyRes}, yields very satisfactory results in general. This observation is not surprising since this testing set shares, by construction, the same feature space properties of the training set, \textsc{Teddy A}. The diagnostics of results, given by the left panel of \tab{tab:Teddy_results}, show that the absolute value of the normalized bias does not exceed $7\times10^{-4}$ and the outlier rate 0.2\% for all four methods. We also see that, due to the color cut, the spec-\emph{z}  of most galaxies of set \textsc{B} is between 0.15 and 0.6.

% Set C - machine learning
% Horizontal line in B and C
Results for \textsc{Teddy C} are shown on the middle panels of \fig{fig:TeddyRes}. While set \textsc{C} shares the same color coverage as set \textsc{A}, the distribution differs. This disparity of color distribution has a minor impact on the photo-\emph{z}scatter. As shown by \tab{tab:Teddy_results}, the std and the outlier rate of \textsc{Teddy C} is slightly higher than that of \textsc{Teddy B}, whereas the mean and the MAD are not affected by a significant change. Readers may notice a horizontal feature privileging a prediction around $\zphoto \approx 0.45$ for both sets \textsc{B} and \textsc{C}. This feature can be explained by the lack of objects in the $r-i$ distribution around 0.7. \fig{fig:colorDistrib_teddy} shows that $r-i$ peaks at 0.6 and 0.8. By examining the colour of these objects, we discovered that photo-\emph{z} predictions for most galaxies with $r-i > 0.7$ are located around $\zphoto\approx0.46$ and predictions for those with $0.6 < r-i < 0.7$ are found around $\zphoto\approx0.36$. Therefore, a local minimum in the galaxy population at $r-i\approx 0.7$ would yield a deficit of predictions around $\zphoto\approx 0.41$, which corresponds to the ``neck'' we observe below the apparent horizontal feature in \fig{fig:TeddyRes}. We conclude that the $r-i$ distributions of sets \textsc{B} and \textsc{C} are responsible for this result.

% Set D - machine learning
% Compare to Sets B & C
Results for \textsc{Teddy D} are shown on the right panels of \fig{fig:TeddyRes} and the left panel of \tab{tab:Teddy_results}. Compared to the two previous cases, the std, MAD, and outlier rate are significantly larger for all methods. As expected, this confirms that if we do not account for the difference of colour coverage between spec-\emph{z}\ and photo-\emph{z}, the assumed error on photo-\emph{z} will be underestimated. 

\begin{figure}
	\includegraphics[width=\columnwidth]{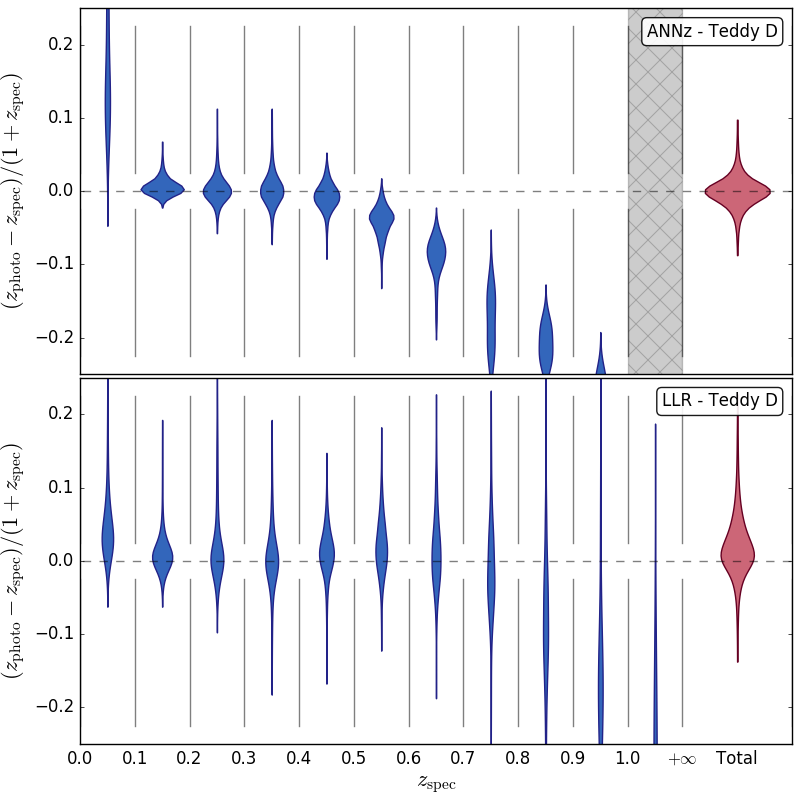}
	\caption{Violin plot of the normalized photo-\emph{z} estimation error on \textsc{Teddy} set \textsc{D}, for the ANNz and LLR methods. The redshift bins have a width of $0.1$.}
    \label{fig:violin_ANNz_LLR_teddy_D}
\end{figure}

% Local vs global
Apart from this general outcome across empirical methods, two distinct behaviours have been observed on set \textsc{D}. The $\zspec$-$\zphoto$ scatter from GAM and LLR stay close to the diagonal, while ANNz and random forest show two horizontal features, resulting in wrong predictions for galaxies with a true redshift at $z<0.2$ or $z>0.45$. Essentially, in the latter case $\zphoto$ values are truncated at the end of the training set coverage. This effect has also been illustrated in \fig{fig:violin_ANNz_LLR_teddy_D}. ANNz clearly has strong $\zspec$-dependent bias, while its overall bias is rather small due to positively and negatively biased regions cancelling out. In contrast, LLR has much smaller redshift-dependent bias, but the overall bias is higher. A similar comparison between random forest and GAM presents almost the same picture, with the exception that GAM also has low overall bias.

%% Violin plot!!!

This result is a consequence of intrinsic differences between the methods we tested. GAM is a method that fits a rather general set of  smooth functions to the training data. There is a ``global'' relationship between covariates and the response variable, which is why we will refer to GAM and similar methods as ``global'' methods. Of course, the function can be evaluated even when there is no coverage of the training set, which is why ``global'' methods are expected to perform well when extrapolating, and indeed that is what we observe on \textsc{Teddy D}.

On the other hand, methods that are strictly depend on the examples present in the training set and do not attempt to extrapolate its features, e.g. nearest neighbours and random forests, are not expected to be able to  perform well beyond the boundaries of the training set -- hence the observed truncation on set \textsc{D}. For such models, the maximum redshift values that can be predicted is determined by the redshifts given in the training data. Neural networks could, in principle, fit an arbitrary ``global'' functional formula, but in practice we observe the same behaviour with ANNz as with the random forest, i.e. dependence dominated by colour-magnitude space neighbours. Therefore we will refer to random forest, ANNz and similar methods as ``local'' methods.

LLR is an interesting hybrid of the previous two classes: it is based on nearest neighbours, thus it should be ``local'', but it also fits a linear functional formula that can be used to extrapolate. Indeed, with enough neighbours ($k=1000$), LLR does extrapolate reasonably well on \textsc{Teddy D}.

% Set B, C, and D - template fitting
\subsubsection{Template fitting methods}

\begin{figure*}
	\centering
	\includegraphics[width=0.95\textwidth]{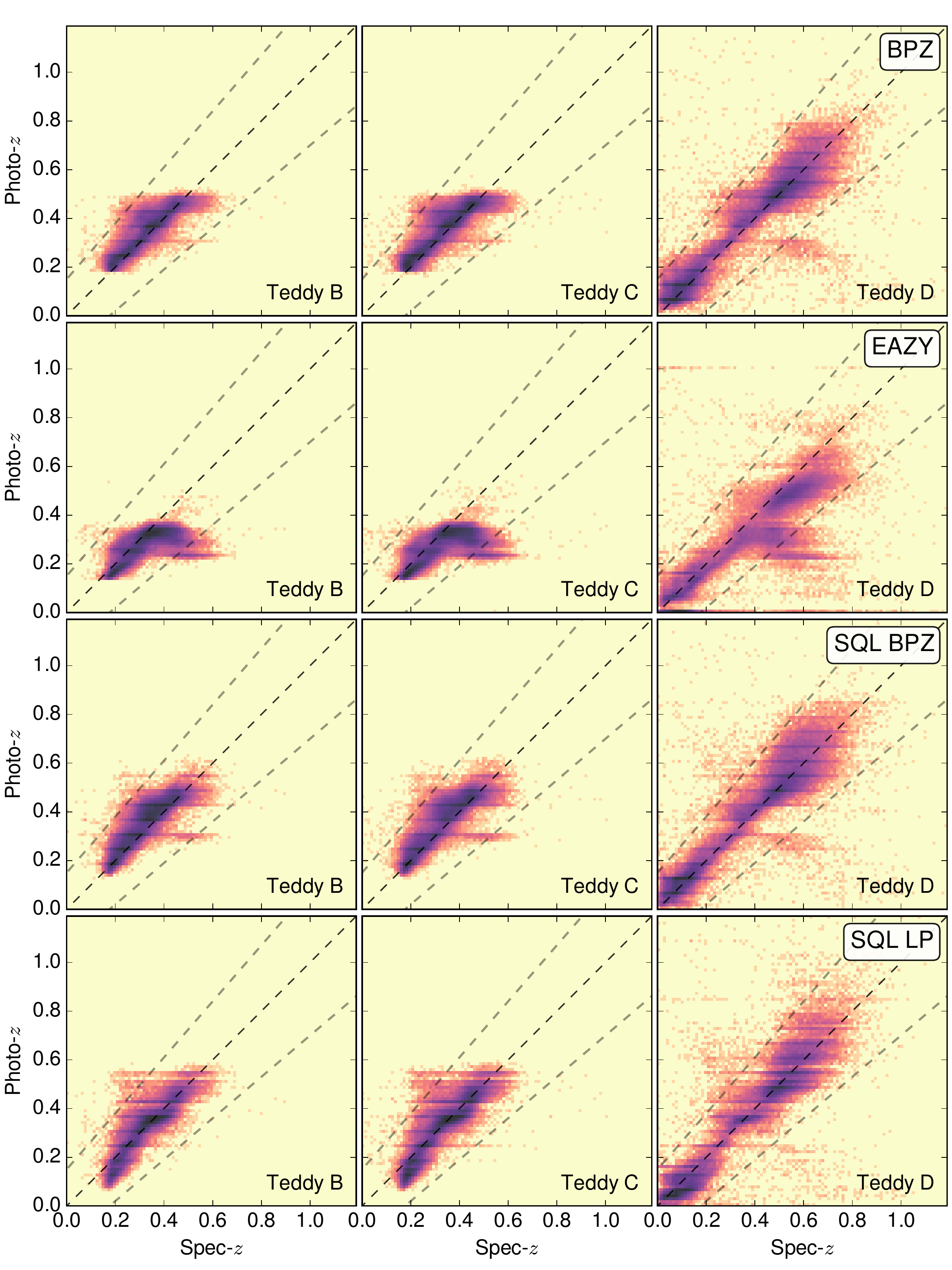}
	\caption{The photo-\emph{z} estimation results for the four template fitting methods (lines) on the three testing subsamples of the \textsc{Teddy} catalogue (columns). The colour gradient shows the logarithmic density. The dashed lines define the perfect prediction (middle) and the limits for being considered as outliers. Numerical results for all 4 diagnostics are shown in \tab{tab:Happy_results} - right panel.}
	\label{fig:scatter_teddy_template}
\end{figure*}

The photo-\emph{z} estimation results on the catalogue \textsc{Teddy} for the four template fitting methods (introduced in \S\ref{sec:methods_template}) are shown in \fig{fig:scatter_teddy_template}, with each row of scatterplots corresponding to a method. Each column represents a subsample of \textsc{Teddy}: in order, sets \textsc{B}, \textsc{C} and \textsc{D}. \textsc{Teddy A} was used as the calibration sample for methods where it was applicable. Numerical diagnostics are presented in the right panel of \tab{tab:Teddy_results}.

The results do not change significantly between \textsc{Teddy B} and \textsc{C} for any of the four methods, only becoming  slightly worse for the latter. The std and MAD values are all in the neighbourhood of $0.03$, and the outlier fraction is $\approx 0.5 \%$ (with the exception of EAZY, where $3-4 \%$). However, most of the bias values are relatively high, comparable to the scatter: $\approx 0.035$ for BPZ and EAZY, and $\approx 0.02$ for SQL BPZ. SQL LP has the lowest bias, $\approx 0.005$. On these samples, the machine learning methods have a clear edge in performance.

A visual inspection of \fig{fig:scatter_teddy_template} reveals that the high bias (and outlier rate) of EAZY is due to a sharp turn away from the diagonal above $\zspec = 0.4$. In contrast, for the two BPZ template methods, the bias is caused by an upward shift of the entire population.

On the sample \textsc{Teddy D}, which is the case illustrating extrapolation outside the coverage of the training set, we can again observe a high bias around $\approx 0.03$ for the BPZ and EAZY methods, with outlier rates of $3.6\%$, $6.6\%$, and std rising to $\approx 0.046$. However, in the case of SQL BPZ and SQL LP, the overall bias remains below $0.01$, with a scatter of $\approx 0.04$ and $2-3 \%$ outliers.

\begin{figure}
	\includegraphics[width=\columnwidth]{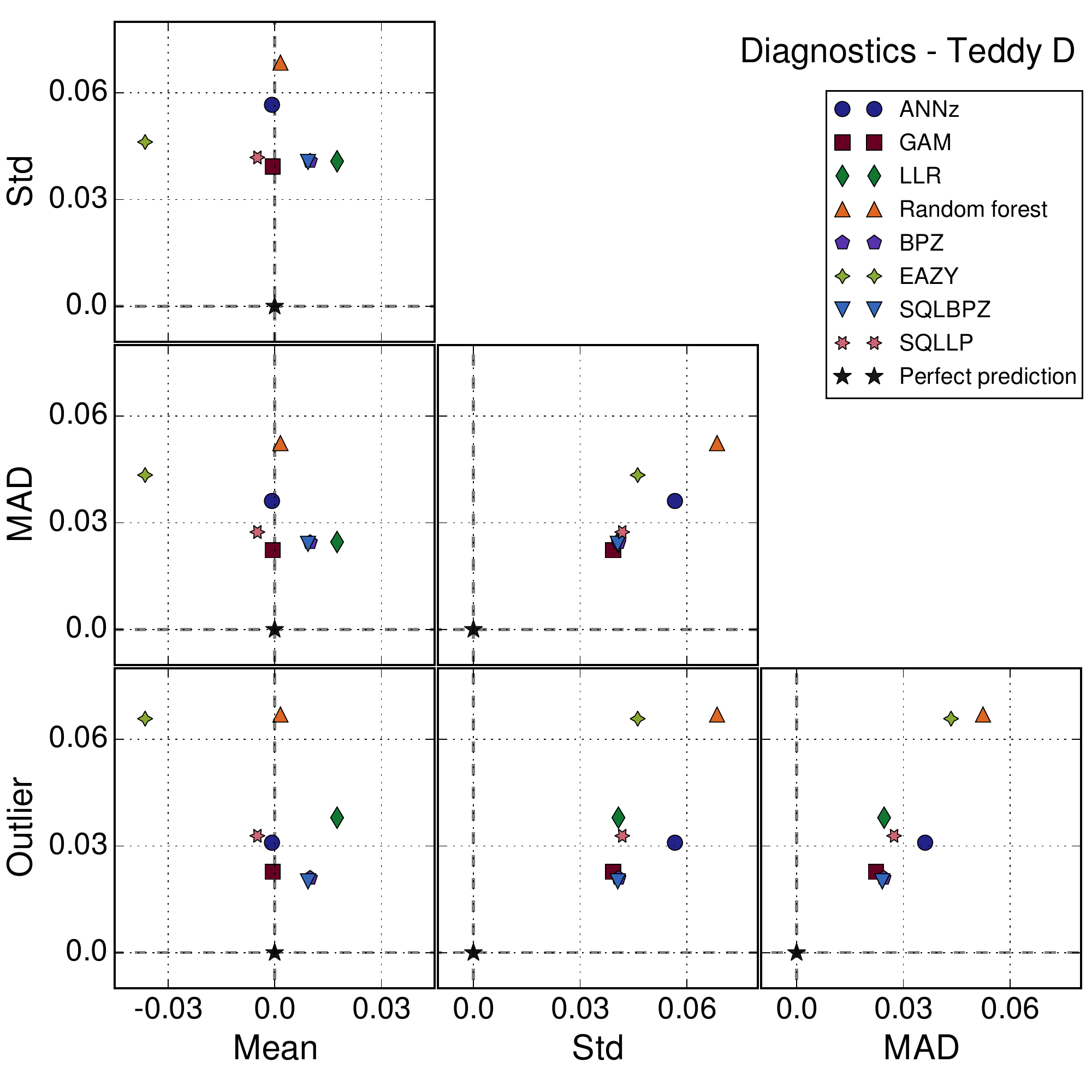}
	\caption{Comparison of numerical diagnostics (panels) for different photo-\emph{z} methods (symbols) from sample \textsc{Teddy D}.} 
    \label{fig:methodComp_teddy_D}
\end{figure}

For the extrapolating case of \textsc{Teddy D}, which is the worst-case scenario in this catalogue, we compare the numerical diagnostics for both empirical and template fitting methods in \fig{fig:methodComp_teddy_D}. In this figure, each photo-\emph{z} method is represented by a different symbol and each panel corresponds to a different diagnostic. The black star in the origin of each panel represents the best-case scenario, where a method would lie. Thus, the closer a symbol is from the black star the better the corresponding method performed according to a given diagnostic. In this visualization, it is possible to note that template fitting methods outperform ``local'', non-extrapolating empirical methods on this sample, and are even comparable to ``global''  methods as long as their overall bias can be managed e.g. with proper calibration. However, GAM still has a slight edge over the best-performing template fitting methods in this test, SQL LP and SQL BPZ.

%%%%%%%%%%%%  Comparison figure  %%%%%%%%%%%%

% Say that teddy is extreme
%\textsc{Teddy} depicts an extreme case because of gap in  $r$-magnitude/colour coverage.  Now we switch to a more realistic scenario, where we shall consider the presence of distinct photometric error distributions and its correlation with the feature space coverage.

%% file: sections/SUBSEC_4.3_happy.tex
\subsection{Results from Happy}

Although \textsc{Teddy} is a good starting point to probe the bias between photometric and spectroscopic samples, it is still simpler than the real scenario. It was intended to isolate the effect of gap in feature space coverage, but also it was built entirely from the SDSS spectroscopic sample, which means that all its objects fulfil the same  spectroscopic data quality requirements. In what follows we shall relax this assumption and quantify the performance of photo-\emph{z} estimators in a harder and more realistic scenario. We now present results for the \textsc{Happy} catalogue -- specially built to account for differences between the photometric error distributions of the samples and their correlation with the lack of feature space coverage.
%The \textsc{Happy} catalogue is intended to showcase how different the photo-\emph{z} estimation performance can be on a realistic photometric sample when compared to validation results on the spectroscopic set.

%\rbeck{TODO Transition, say that Happy is more realistic, and that we needed better quality external spectroscopic measurements to realize this validation (Teddy can't cover the errors of Happy).}

\subsubsection{Machine learning methods}

\label{sec:happy_ML}

\begin{figure*}
	\centering
	\includegraphics[width=0.95\textwidth]{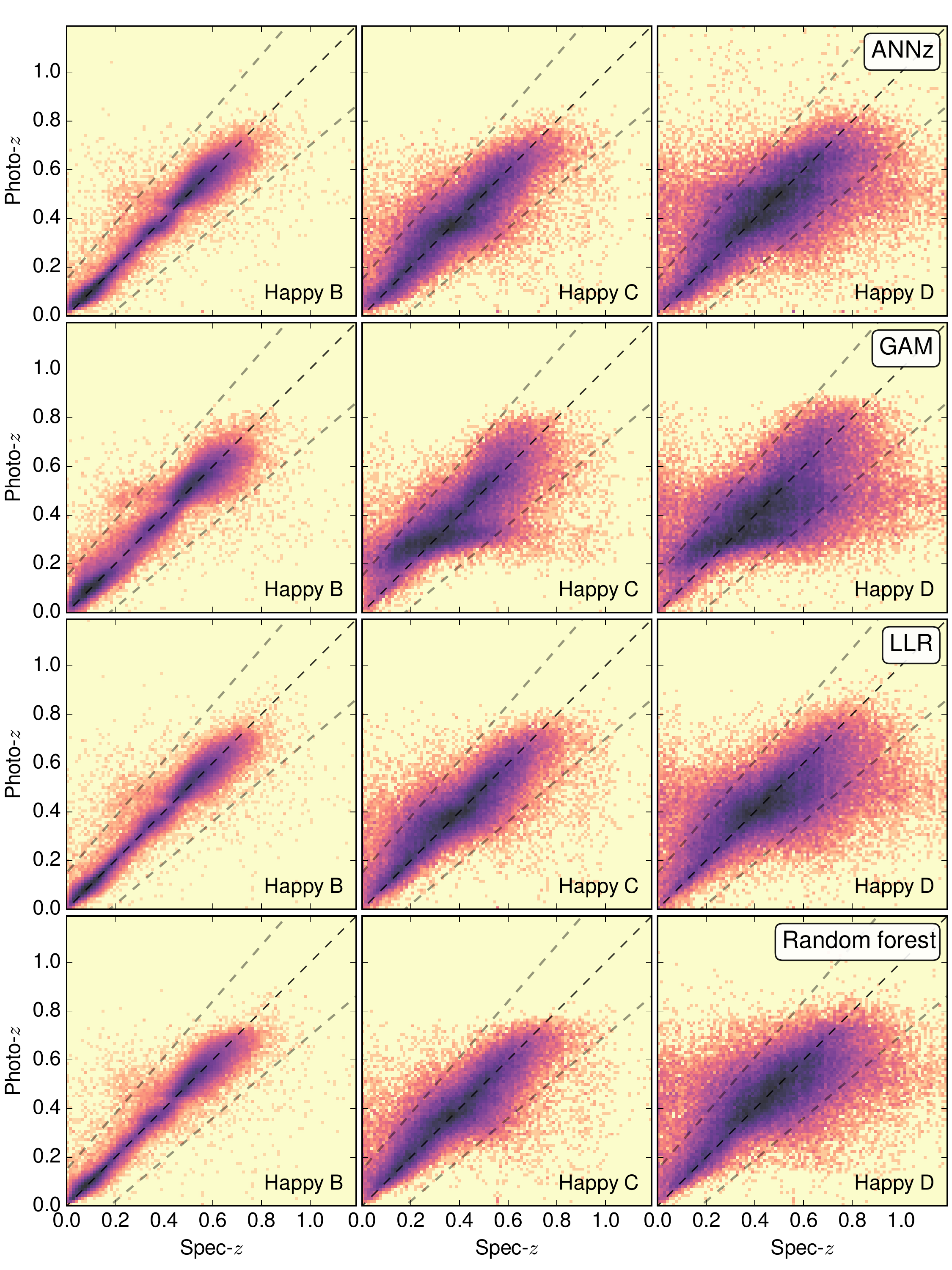}
	\caption{Results from applying empirical photo-\emph{z} algorithms (lines) to the 3 testing samples of the \textsc{Happy} catalogue (columns). The colour gradient shows the logarithmic density. The dashed lines define the perfect prediction (middle) and the limits for being considered as outliers. Numerical results for all 4 diagnostics are shown in \tab{tab:Happy_results} - left panel.}
	\label{fig:scatter_happy_ML}
\end{figure*}

\fig{fig:scatter_happy_ML} shows the scatterplots of photo-\emph{z} estimation results for the machine learning methods described in \S\ref{sec:methods_empirical} on different samples of the \textsc{Happy} catalogue. %Again, each column corresponds to a different subsample: left, center and right to \textsc{Happy B}, \textsc{C} and \textsc{D}, respectively. Each row depicts the results of a given method, indicated in the top right corner of the rows. 
Numerical diagnostics are presented in the left panel of Table~\ref{tab:Happy_results}.

%%%%%%%%%%%%  Happy table  %%%%%%%%%%%%
\begin{table*} 
	\centering
    \begin{minipage}{0.49\textwidth}
	\centering
	\begin{tabular}{cccccc}
    	\hline\hline\\[-2.25ex]
		\multirow{3}{*}{Method} & \multirow{3}{*}{Set} & \multicolumn{4}{c}{Diagnostics}\\[0.5ex]
 		&  & Mean & Std & MAD & Outlier rate\\ 
        &  & ($\times 10^{-2}$) & ($\times 10^{-2}$) & ($\times 10^{-2}$) & (\%) \\
        \cline{2-6}
		\hline \noalign{\vspace{0.1cm}}
		\multirow{3}{*}{ANNz}  & B & 0.04 & 2.87 & 1.49 & 0.99\\ 
                               & C & 0.16 & 5.41 & 3.60 & 5.59\\ 
                               & D & -0.52 & 6.53 & 5.44 & 14.01\\ 
       \hline \noalign{\vspace{0.1cm}} 
       \multirow{3}{*}{GAM}  & B & 0.09 & 3.50 & 1.95 & 1.36\\ 
                             & C & 0.86 & 6.34 & 4.84 & 7.37\\ 
                             & D & -0.51 & 7.21 & 6.70 & 16.38\\ 
       \hline \noalign{\vspace{0.1cm}} 
       \multirow{3}{*}{LLR}  & B & 0.13 & 2.81 & 1.39 & 1.11\\ 
                             & C & 0.52 & 5.45 & 3.59 & 6.07\\ 
                             & D & -0.79 & 6.62 & 5.62 & 14.52\\ 
       \hline \noalign{\vspace{0.1cm}} 
       \multirow{3}{*}{\parbox{1.1cm}{\centering Random\\Forest}}   & B & 0.05 & 2.82 & 1.41 & 1.02\\ 
                                    & C & 0.34 & 5.39 & 3.51 & 5.58\\ 
                                    & D & -0.28 & 6.51 & 5.36 & 14.2\\
        \hline
	\end{tabular}
	\end{minipage}
    \begin{minipage}{0.49\textwidth}
	\centering
	\begin{tabular}{|cccccc}
    	\hline\hline\\[-2.25ex]
		\multirow{3}{*}{Method} & \multirow{3}{*}{Set} & \multicolumn{4}{c}{Diagnostics}\\[0.5ex]
 		&  & Mean & Std & MAD & Outlier rate\\ 
        &  & ($\times 10^{-2}$) & ($\times 10^{-2}$) & ($\times 10^{-2}$) & (\%) \\ 
		\hline \noalign{\vspace{0.1cm}} 
		\multirow{3}{*}{BPZ}  & B & 2.11 & 3.93 & 2.81 & 1.88\\ 
                              & C & 0.21 & 5.81 & 4.20 & 7.97\\ 
                              & D & -1.56 & 6.66 & 6.41 & 20.1\\ 
       \hline \noalign{\vspace{0.1cm}} 
       \multirow{3}{*}{EAZY}  & B & -3.66 & 4.57 & 4.27 & 6.31\\ 
                              & C & -4.11 & 5.48 & 6.25 & 17.88\\ 
                              & D & -4.23 & 6.19 & 9.17 & 31.72\\ 
      \hline \noalign{\vspace{0.1cm}} 
      \multirow{3}{*}{SQL BPZ}  & B & 1.79 & 4.12 & 2.75 & 1.80\\ 
                               & C & 0.09 & 5.94 & 4.41 & 8.87\\ 
                               & D & -1.82 & 6.77 & 6.80 & 21.25\\ 
      \hline \noalign{\vspace{0.1cm}} 
      \multirow{3}{*}{SQL LP}  & B & -0.47 & 4.15 & 2.68 & 3.20\\ 
                              & C & -0.51 & 5.90 & 4.61 & 14.18\\ 
                              & D & -1.33 & 6.74 & 8.63 & 34.14\\ 
     \hline
	\end{tabular}
	\end{minipage}
	\caption{Results for the \textsc{Happy} catalogue.}
	\label{tab:Happy_results}
\end{table*}
%%%%%%%%%%%%%%%%%%%%%%%%%%%%%%%%%%%%%%%%%%%%%%%%%%%%%%

All methods provide reasonably accurate redshifts on \textsc{Happy B}, where the estimated galaxies have the same distribution of magnitude, colour \textit{and} photometric error as the spectroscopic training set (\textsc{Happy A}). Outlier rates were kept around $1\%$ for all tested photo-\emph{z} codes while local methods (ANNz and random forest) presented smaller biases, $\approx 5 \times 10^{-4}$, then global ones (LLR and GAM), $\approx 1 \times 10^{-3}$. GAM obtained the larger scatter $\approx 3.5 \times 10^{-2}$, a trend that was maintained for the other samples. 

Interestingly, even with the same range of photometric errors, and within the coverage of other properties, on \textsc{Happy C} the scatter and proportion of outliers are significantly larger across the board due to the different error distribution (weighted towards higher errors).
In this context, ANNz obtained significantly smaller bias than the other methods, $0.16 \times 10^-2$ - half of the number achieved by random forest, the second largest biased - and GAM presented the largest outlier rates, $\approx 7 \%$ and scatter, $\approx 6.3$. As expected, photo-\emph{z} accuracy drops even more for all methods on \textsc{Happy D}, where many objects are outside the coverage of the training set in all respects. In this scenario, all methods produced outlier rates $> 10\%$.
For the GAM method, an unwanted feature shows up on \textsc{Happy C}, a linear broadening of the well-populated region between $\zspec \approx 0.2-0.6$ that is estimated to be at $\zphoto \approx 0.3$. The feature broadens even further on the sample \textsc{Happy D}. This would suggest that the global fitting is more and more disrupted as the photometric errors increase, whereas for the other local methods such an effect is not observed. The numerical results also show that indeed GAM performs worse than the other methods when photometric errors get higher: the MAD value goes from being $\approx 0.005$ worse than the other methods to being $\approx 0.012$ worse, while the outlier rate goes from $ 0.3 \%$ worse to $ 2 \%$ worse.

There are two main takeaways from the results on the \textsc{Happy} catalogue.
First, even if an error or colour cut is performed on the photometric sample to make it cover the same parameter range as the spectroscopic training set (this was done in \textsc{Happy C}), the results on a spectroscopic validation set (\textsc{Happy B}) will not be representative of results on such a cut photometric sample, due to the differing shape of the error distribution. Ultimately, to accurately determine the photo-\emph{z} estimation accuracy of an object, its individual photometric error has to be taken into account along with the typical photo-\emph{z} error of its NN galaxies (those with similar colour and magnitude).
It follows that any attempt at dealing with the mismatch between spectroscopic and photometric samples must also include their photometric error distribution differences in the calculation of population diagnostics, independently of their feature space coverage. Otherwise, even adaptations like calculating appropriate weights for the training sample will output too optimistic diagnostics (see \S\ref{subsec:DA}). The \textsc{Happy} catalogue provides for the first time an environment where such new approaches can be directly tested. %\rbeck{e.g. by comparing diagnostics on \textsc{Happy C} without adaptation, and results on \textsc{Happy B} with adaptation turned on.}

Second, based on the methods tested here, it appears that global model fitting methods perform worse in the presence of large photometric errors than local empirical methods. Neural networks, while in essence fitting an arbitrary functional formula, behave similarly to nearest neighbour methods in this regard.

\subsubsection{Template fitting methods}

\begin{figure*}
	\centering
	\includegraphics[width=0.95\textwidth]{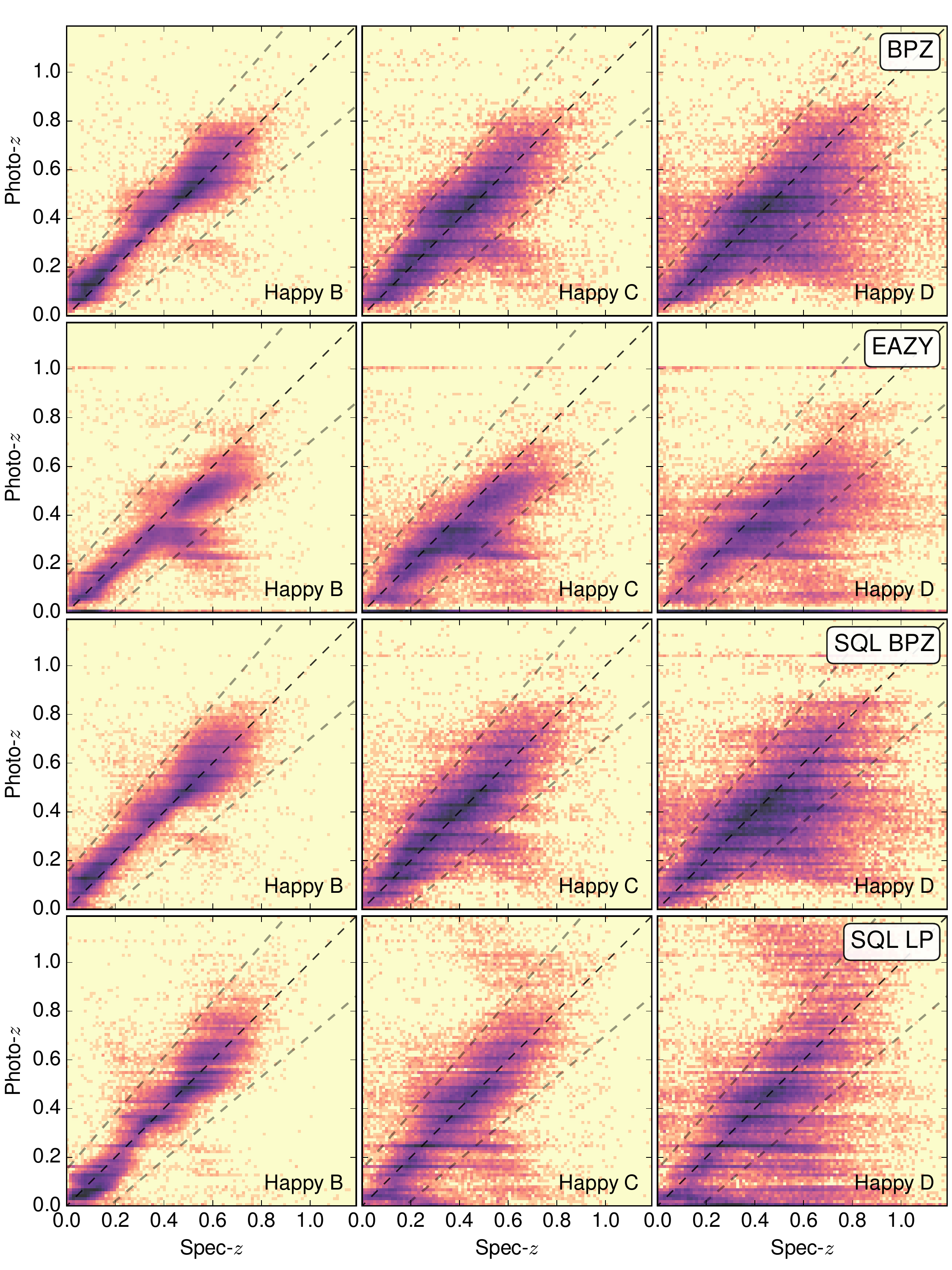}
	\caption{Template fitting photo-\emph{z} results obtained from the 4 methods described in \S\ref{sec:methods_template} (lines) for the three testing subsamples of the \textsc{Happy} catalogue (columns). The colour gradient shows the logarithmic density. The dashed lines define the perfect prediction (middle) and the limits for being considered as outliers. Numerical results for all 4 diagnostics are shown in \tab{tab:Happy_results} - right panel.}
	\label{fig:scatter_happy_template}
\end{figure*}

The template fitting based photo-\emph{z} estimation scatterplots for the \textsc{Happy} catalogue are presented in \fig{fig:scatter_happy_template}, using the same layout as in previous such figures. Numerical diagnostics are shown in the right panel of Table~\ref{tab:Happy_results}.

For the \textsc{Happy B} sample, all methods perform reasonably well, but with a notable overall positive bias of $\approx 0.02$ for the two BPZ template cases, and a negative bias of $\approx - 0.035$ for EAZY. The Le PHARE (LP) template case had the least bias, roughly $- 0.005$. On \textsc{Happy C}, with the photometric errors increasing, the scatter also jumps, but the degraded performance is best illustrated by the outlier rate skyrocketing to $8-18\%$. On \textsc{Happy D} this trend only continues, with outlier rates becoming unmanageable, between $20-34\%$.

We note that the many extreme outliers of the SQL LP case are a result of overfitting the errors -- the Le PHARE template set is rather varied ($641$ templates in this configuration), containing young, dusty starburst galaxies with different dust models, thus more extreme colour combinations can still be fitted. Using a prior, the number of extreme cases could have been greatly mitigated, but that would have increased bias on \textsc{Happy B}, which is the sample we chose to optimise for.

An interesting feature, a populated outlying region appears around $\zspec =0.5$ and $\zphoto = 0.3$ for most cases in \fig{fig:scatter_happy_template}, potentially indicating a systematic effect in the SDSS measurements that leads to erroneous template matches. It also mostly coincides with the elongated linear feature of the GAM method on \textsc{Happy C} (see \S\ref{sec:happy_ML}).

%\emi{for Robert:
%\begin{itemize}
%\item impact of photometric errors. 
%\item how photometric errors are expected to improve in future surveys 
%\item discuss if it will be enough to solve this problem (we hope not). 
%\end{itemize}}

%\rbeck{I would stick to the results here, and put these thoughts into the discussion. I don't think we can give an answer to this based on our results, anyway.}

\begin{figure}
	\includegraphics[width=\columnwidth]{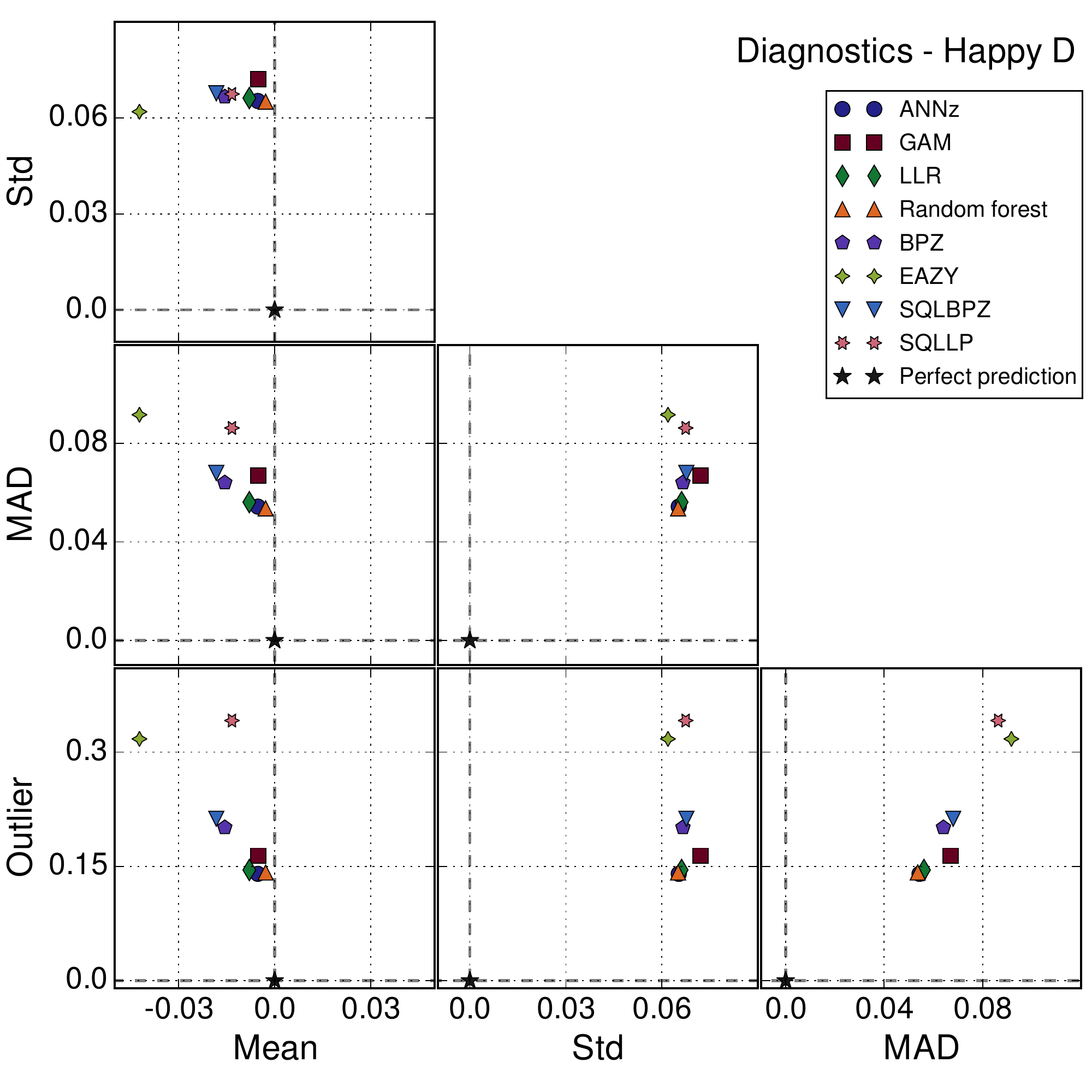}
	\caption{Comparison of numerical diagnostics  obtained from applying different photo-\emph{z} methods to \textsc{Happy D}.} 
    \label{fig:methodComp_happy_D}
\end{figure}

%\rbeck{Should we start a new subsubsection for the following short part?}
A summary of the diagnostics for the most realistic case, depicted in \textsc{Happy D}, is shown in 
%The performance of all the methods tested -- as probed by the numerical diagnostics -- is compared in 
\fig{fig:methodComp_happy_D}. The configuration of the plot is the same as shown in \fig{fig:methodComp_teddy_D}. The machine learning methods all have similar performance, with GAM being slightly worse in terms of std, and the template fitting methods clearly lagging behind. However, we note that even the results of the best-performing method leave much to be desired.

%%%%%%%%%%%%  Comparison figure  %%%%%%%%%%%%

%% file: sections/SUBSEC_4.4_DA.tex
\subsection{Reshaping photometric feature space distributions}
%\subsection{Density and ratio estimation}
\label{subsec:DA}

The results presented above demonstrate how \textsc{Teddy} and \textsc{Happy} allow us to quantify the impact of spectroscopic coverage and photometric errors in photo-\emph{z} methods, respectively. In this section we show how they can also provide an environment to assess the efficiency of possible solutions. Consider the case of different probability distributions in feature space as depicted  in the \textsc{C} samples from both  catalogues. In such cases, when distributions in feature space between the spectroscopic and photometric samples are significantly different, the mismatch may be incorporated into the learning scheme.  In the machine learning community such strategies are often classified as \emph{Domain Adaptation} (DA) techniques \citep{Quionero2009}. 

%case where the spectroscopic (training) sample covers all the range occupied by the photometric (test) sample, in feature space, it is possible to minimize the effects of statistically significant differences in distribution by giving appropriate weights to the objects in the spectroscopic sample. 
%the distributions in feature space between spectroscopic confirmed and photometric objects might different significantly. The general idea of covariate shift is to compensate this mismatch between both set by considering the differences of the two distributions. 
%A direct approach to this problem is to estimate both distributions and to incorporate the mismatch between them into the learning scheme. 
\emph{Importance weighting}~\citep{HuangSGBS2007} represents  one possible domain adaptation approach for the learning scenarios at hand. It states that it is possible to reweight the training examples in order to increase the importance of entries that are frequent in the test, but under-represented in the training sample. This is achieved by assigning higher weights to such test instances. In a similar fashion, samples that are over-represented in the training set are downweighted. Hence, such a reweighting strategy aims at reducing the shift between training and testing distribution in the feature space. % (i.e., w.r.t. a shift in $p(\mathbf{x})$). [REMOVE THE NEXT SENTENCE, KEEP PREVIOUS NEW ONE?: This strategy is based on the hypothesis that two samples with identical distributions in feature space will also have identical distributions in redshift, $N(z)$].} 
In astronomy, an implementation of similar ideas was presented by \citet{Lima2008} and has been applied to large imaging surveys \citep[e.g.][]{bonnett2016}, while other forms of domain adaptation were also applied to star classification problems \citep[e.g.][]{vilalta2013}. 

\citet{HuangSGBS2007} also propose a so-called \emph{kernel mean matching} (KMM) framework that aims at estimating corresponding weights via quadratic programming. This framework has been applied to the problem of supernova photometric classification leading to encouraging results \citep{pamplana2016}. Albeit technically robust, the approach is limited by the amount of both training and test points it can handle. For the scenarios considered in this work, the data sets can easily consist of hundreds of thousands of objects -- too large for standard quadratic programming solvers. For such cases, \cite{Kremer2015} have extended a nearest neighbour based technique that scales well for large samples, especially given low-dimensional search spaces. We used this machinery\footnote{Code available at \href{https://github.com/kremerj/nnratio}{https://github.com/kremerj/nnratio}.}, with the default configuration to calculate the weights for objects in samples A in all the 3 different scenarios with test samples B, C and D considering both our catalogues. The reader should be aware that the application of this method in sparse test samples, such as those in \textsc{Happy}, might lead to numerical problems. In order to ensure the convergence of the results we constructed larger test samples, following the same procedure described in \S\ref{sec:design:happy}, for the single purpose of assuring the convergence of the weight coefficients. The extended \textsc{Happy C} catalogue is also publicly available. 

Once the weights were calculated, we incorporated them into the GAM and LLR methods, which are the ones allowing an easy implementation of these coefficients without the need of complex modifications in the original code. 
As expected, results considering  sets B as test samples yield the same outcomes as presented in the left panels of  Figures \ref{fig:TeddyRes} and \ref{fig:scatter_happy_ML}. %Given that B sets  are completely representative of  A, there were no significant differences between the weights assigned to the training objects in this situation.  
Similarly, results from sets D  were not that different from the original case. They were built to emphasize the gap in coverage between training and test sets -- a situation where DA is not expected to have a large effect. As stated before, supervised learning algorithms learn by example and their results should not be extrapolated beyond the range covered by the training sample. 

Samples \textsc{C} on the other hand, are a good testing ground, since in their case 
%Interesting outcomes were obtained from \textsc{Happpy} C, where 
the spectroscopic sample provides at least a few examples in all regions of the parameter space occupied by the photometric sample. Using GAM in \textsc{Teddy C} we obtained a bias of $3 \times 10^{-4}$ - half of the value achieved without considering the weights  (see \tab{tab:Teddy_results}), while other diagnostics were not significantly changed. On the other hand, LLR did not show noticeable deviations between the weighted and unweighted results. This was also expected given the local characteristic of the method. Since LLR performs the regression exercise considering only a certain region of the parameter space, the weights calculated based on density estimates were not much different for a given set of neighbours.

\begin{figure}
	\centering
	\includegraphics[width=\columnwidth]{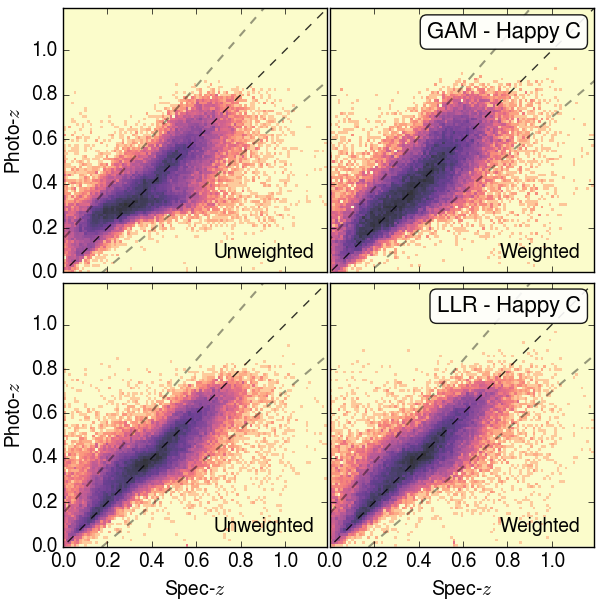}
	\caption{Comparison between results from LLR and GAM methods in the traditional approach (left) and after re-weighting  the spectroscopic sample (right) for set \textsc{Happy} C. }
	\label{fig:scatter_weight}
\end{figure}

Interesting qualitative results were encountered in \textsc{Happy C}. The weighting method was able to remove the horizontal trend mentioned in \S\ref{sec:happy_ML}, resulting in a much more homogeneous distribution in $\zspec \times \zphoto$ space (see \fig{fig:scatter_weight}), although numerical diagnostics were not significantly changed. 
%In particular, \textsc{Happy} C provided some interesting results. Figure \ref{fig:scatter_weight} shows how the method is able to homogenise the bias in intermediate redshifts we detected before for GAM. 
The latter is a direct consequence of the different underlying error distributions between training and test samples, as noted in \S\ref{sec:happy_ML}. The horizontal feature on the figure, which was associated with different underlying photometric feature distributions, was minimized, but in order to make a difference in the population diagnostics it is imperative to also take the error distributions into account.

 %Such strategies might help to improve the performance of certain models as indicated by the results provided above. 
%From a real-world perspective, there are typically other shifts as well in astronomical catalogs such as shifts w.r.t. the target distribution $p(y)$ (e.g., due galaxies with a higher redshift being more interesting to the community) or shifts in $p(y|\mathbf{x})$ (i.e., given a region in the feature space, the distribution in $y$ might still vary from training to test data). Dealing with such additional shifts depicts a very challenging topic.

%%%%%%%%%%%%%%%%%%%%%%%%%%%%%%%%%%%%%%%%%%%%%%%%%%%%%%%%%%%%%%%%%%%%%%%%%%%%%%%%

%% file: sections/SEC5_discussion.tex
\section{Discussion}
\label{sec:disc}

% what we did in this paper
In this paper we present a comprehensive discussion of two main differences between spectroscopic and photometric samples which must be addressed by photo-\emph{z}  analyses: the  lack of complete spectroscopic coverage in the colour/magnitude space and the presence of distinct photometric error distributions and ranges. 

%of the distribution in colour and magnitude space between spectroscopic and photometric samples and the effect of different photometric error distribution and range  the photo-\emph{z} performance

These problems are well known within the photo-\emph{z} community but up to now there was no standard environment enabling a quantitative analysis of their influence on the final photo-\emph{z} estimates. We developed such an environment through the construction of 2 catalogues: \textsc{Teddy} aims at isolating the effect of incomplete spectroscopic coverage and \textsc{Happy} probes the effect of photometric error distributions and ranges. Both catalogues are composed of 4 samples, with sample \textsc{A} being representative of a spectroscopic sample - and thus should be used for training purposes - and samples \textsc{B, C} and \textsc{D} depicting increasingly complex data situations. \textsc{Teddy} was built completely from the SDSS-DR12 spectroscopic sample while \textsc{Happy} uses photometry from SDSS and spectroscopy from many different sources (\S\ref{sec:design}) -- this allowed us to reproduce the statistical properties of the real SDSS-DR12 photometric sample while still possessing spectroscopic measurements for all objects in \textsc{Happy} samples. Both catalogues were submitted to the scrutiny of various machine learning and template fitting photo-\emph{z} methods -- both established and new, such as \textit{Generalized Additive Models} (GAM).

%First, using our new catalogue \textsc{Teddy}, we investigated how the mismatch of the distribution in colour and magnitude space between spectroscopic and photometric samples 
%affects photo-\emph{z} results. Second, with the   \textsc{Happy} catalogue, we examined how different photometric error distribution and range degrade the photo-\emph{z} performance.

% results from applying the methods - with coverage
 We confirmed that most methods can adequately handle a difference in distribution shape as long as there is sufficient coverage in the training sample. In this set-up, template fitting methods perform worse, especially in terms of bias. 
% results from applying the methods - no coverage
However, when training set coverage is not available and thus extrapolation must be performed, local machine learning methods fail, while global methods achieve reasonable results (see \S\ref{sec:teddy_ML}). Also, in the extrapolation case template fitting methods with the proper configuration can perform comparably to the better machine learning methods.
% comment on z extrapolation on teddy
It should be noted that our extrapolation test sample (\textsc{Teddy D}) still has a relatively large number of training set points, and although their colour coverage is rather limited, we do not have to extrapolate very far in terms of redshift. More extensive extrapolation might prove to be difficult for all kinds of machine learning methods. %machine learning methods.

% cutting the colour distribution is not enough
We demonstrated that even with a photometric error cut in the photometric sample, a differing error distribution can lead to significantly worse results than what is observed on a traditional spectroscopic validation set. Increasing measurement error impairs performance for all methods, with template fitting methods being affected the most, global machine learning methods behaving better, and finally local ML algorithms proving the most error-resistant.
% how to choose a photoz method
Ultimately, the choice of photo-\emph{z} method in applications will have to be optimised for the data at hand, specifically the colour-magnitude \textit{and} error distributions of both the available training set and the target photometric sample. When choosing between various template fitting and machine learning methods, a trade-off has to be made between extrapolation capability, performance within the coverage of the training set, and error resistance.
% cannot ignore photometric errors
With advances in instrumentation, the more accurate photometry of upcoming surveys \citep{Stubbs2007PASP,Li2016AJ} and the relative lack of corresponding extensive spectroscopic samples might well tip the focus of such photo-\emph{z} method optimisation towards extrapolation, favouring template fitting or global machine learning methods. However, ground-based observations do have strong physical limits regarding achievable photometric accuracy \citep{Hartman2005,Stubbs2007PASP}, therefore the error resistance property cannot be ignored, either.

In order to illustrate how the catalogues presented here can be used to test alternative techniques aimed at dealing with the issues described above, we apply a density ratio weighting of the training sample which is designed to deal with differences in feature distributions (\S\ref{subsec:DA}). Our results show that the method is able to deal with biases in photo-\emph{z} determination but it does not numerically improve the overall diagnostics. This was expected, since such reweighting schemes can only reduce potential differences in feature space distribution, and cannot help with coverage or error distribution. %Such strategies might help to improve the performance of certain models as indicated by the results provided above. 
In our context, the most realistic scenario with adequate coverage, depicted in \textsc{Happy C}, has a distinct photometric error distribution which was not taken into account. This is a clear example that simply reweighting the training sample is not enough to provide a realistic estimation of photo-\emph{z} accuracy. Also, from a real-world perspective, in astronomical catalogues there could be shifts other than the ones mentioned so far, e.g. shifts in conditional probability distributions. Dealing with such additional shifts is a serious challenge yet to be overcome.

%e.g., due galaxies with a higher redshift being more interesting to the community or shifts in $p(y|\mathbf{x})$ (i.e., given a region in the feature space, the distribution in $y$ might still vary from training to test data). Dealing with such additional shifts depicts a very challenging topic. \rbeck{TODO review this - maybe first mention DA in the Discussion, and then this can follow?}

% why we did not use quality cuts
We note that most methods have quality cuts that could be used to filter out the worst photo-\emph{z} failures. For example, template fitting methods can recognise bad or improbable template matches, and wide or multi-peaked posterior redshift PDFs, while machine learning methods can signal extrapolation, large nearest neighbour bounding boxes, large deviations among neighbours, or sparse training set coverage. However, we saw no way to make the same quality cut across all the methods tested here, therefore in the spirit of fairness we chose not to do \textit{no} quality cuts. Moreover, our goal is to emphasize that results grow worse through samples \textsc{B} to \textsc{D}, and that unless the issues addressed here are accounted for, we cannot explore the full potential of our data sets. 
%The two catalogues we created to explore these issues, \textsc{Teddy} and \textsc{Happy} could be useful test benches in proving the fitness of current and future photo-\emph{z} approaches. We strongly recommend using such difficult test cases as opposed to a simple cross-validation within a spectroscopic sample, since the latter was shown not to give representative results for the realistic use case.

% explain the titl
Finally, the fact that external, better quality spectroscopy was needed for \textsc{Happy} to properly mimic the real photometric sample characteristics while also possessing spectroscopic measurements is crucial -- when one only has spectroscopic observations from the same instrument, it is impossible to test results on the poorer quality photometric observations. Thus, \textsc{Teddy} will never be \textsc{Happy}, or in other words, it is not possible to realistically address the photo-\emph{z} challenges with only the uniform spectroscopic-quality measurements of a given survey, and without taking more than the feature space coverage into account.

%and the additional testing provided by our catalogues becomes essential to ensure the full exploitation of future large scale surveys.

% what do we hope with this?
\textsc{Teddy} and \textsc{Happy} are publicly available\footnote{\href{https://github.com/COINtoolbox/photoz_catalogues}{https://github.com/COINtoolbox/photoz\_catalogues}}. They were built to be used as test benches in proving the fitness of current and future photo-\emph{z} approaches. We strongly recommend using such difficult test cases as opposed to a simple cross-validation within a spectroscopic sample, since the latter was shown not to give representative results for the realistic use case. We also hope that providing such a user-friendly environment to test these issues will encourage not only astronomers, but researchers from other related areas to approach this problem.